\documentclass[aps,pra,reprint,superscriptaddress]{revtex4-2}
\usepackage{graphicx}
\usepackage{dcolumn}
\usepackage{bm}
\usepackage{wasysym}
\usepackage{amssymb}
\usepackage[breaklinks]{hyperref}
\hypersetup{colorlinks=true,breaklinks=true,urlcolor=blue,linkcolor=blue,citecolor=blue}

\usepackage[dvipsnames]{xcolor}

\begin{document}


\title{Comprehensive characterization of an apparatus for cold electromagnetic dysprosium dipoles}

\author {Gregor Anich}
\affiliation {Institut für Quantenoptik und Quanteninformation, Österreichische Akademie der Wissenschaften, 6020 Innsbruck, Austria}
\affiliation {Institut für Experimentalphysik, Universität Innsbruck, 6020 Innsbruck, Austria}
\author {Niclas H\"{o}llrigl}
\affiliation {Institut für Quantenoptik und Quanteninformation, Österreichische Akademie der Wissenschaften, 6020 Innsbruck, Austria}
\affiliation {Institut für Experimentalphysik, Universität Innsbruck, 6020 Innsbruck, Austria}

\author {Marian Kreyer}
\affiliation {Institut für Quantenoptik und Quanteninformation, Österreichische Akademie der Wissenschaften, 6020 Innsbruck, Austria}
\author {Rudolf Grimm}
\affiliation {Institut für Quantenoptik und Quanteninformation, Österreichische Akademie der Wissenschaften, 6020 Innsbruck, Austria}
\affiliation {Institut für Experimentalphysik, Universität Innsbruck, 6020 Innsbruck, Austria}


\author{Emil Kirilov}
\affiliation {Institut für Quantenoptik und Quanteninformation, Österreichische Akademie der Wissenschaften, 6020 Innsbruck, Austria}
\affiliation {Institut für Experimentalphysik, Universität Innsbruck, 6020 Innsbruck, Austria}



\date{\today}

\begin{abstract}
We report on the development of an advanced ultracold dysprosium apparatus, which incorporates a cold atom microscope (CAM) with a design resolution of a quarter micrometer. The CAM and the cooling and trapping regions are within the same vacuum glass vessel ensuring simple atom transport between them. We demonstrate the essential experimental steps of laser and evaporative cooling, lattice loading,  transporting and precise positioning of a cloud of the bosonic isotope $^{164}$Dy at the CAM focal plane. Basic characterization of the CAM and future plans in enabling its full capacity are outlined. 
We also present a feasible platform for simulating complex spin models of quantum magnetism, such as the $XYZ$ model, by exploiting a set of closely spaced opposite parity levels in Dy with a large magnetic and electric dipole moment. We isolate a degenerate isospin-1/2 system, which possesses both magnetic and electric dipole-dipole coupling, containing Ising, exchange and spin-orbit terms. The last gives rise to a spin model with asymmetric tunable rates that depend on the lattice geometry.  
\end{abstract}


\maketitle

\section{Introduction}
Low-temperature systems with dominant dipole-dipole interaction (DDI), which is both long range and anisotropic~\cite{Baranov2012}, have recently sparked an enormous interest because of intriguing prospects related to the simulation of condensed matter systems, quantum information and ultracold chemistry~\cite{Carr2009cau}.
A simple example already containing remarkably complex physics is a system of free dipoles in a 2D geometry~\cite{Buechler2007brief}, with a direct resemblance to the behavior of a 2D electron gas in metal-oxide-semiconductor field-effect transistors~\cite{Spivak2004Oct}. At higher temperatures the system presents a unique opportunity to study the Berezinskii-Kosterlitz-Thouless (BKT) transition, expected to take place between a normal fluid and superfluid~\cite{Buechler2007brief}. Another unique phase, coined  the micro-emulsion phase~\cite{Spivak2004Oct}, should mediate the transition between a Fermi liquid and a Wigner crystal at low temperatures.

Models of quantum magnetism are paradigms of key importance in strongly correlated many-body physics~\cite{Perrin2022}. The theoretical outset when one tries to recast those models with cold dipolar particles, occupying the sites of an optical lattice, lies in the projection of the DDI operator onto a few isolated magnetic sublevels. This results in spin-spin coupling, which can simulate a plethora of lattice spin models utilized to describe quantum magnetism, such as the generalized $XXZ$~\cite{Jafari2007}, generalized $XYZ$ model~\cite{Fernanda2013, Wall2015Feb}, or models of high-$T_\mathrm{c}$ superconductivity~\cite{Ogata2008} like the $t$-$J$ model~\cite{Chao1977May}. 

To give an instance, some of these experimental platforms that utilize DDI between particles allow for precise tuning of the mutual weights of the spin exchange and Ising terms comprising the $XXZ$ Hamiltonian. This is essential in order to map out the full phase diagram, and possibly observe the expected phase transition from paramagnetic spin fluid to N\'eel-ordered antiferromagnet. The possibility to quench the system across the different regions of its phase diagram should spontaneously break global symmetries and therefore enable the Kibble-Zurek mechanism~\cite{Kibble1976,Zurek1985}. 

Furthermore of particular interest are magnetic systems with implemented competing interactions that cannot be fulfilled simultaneously, and therefore generate frustration~\cite{Balents2010}. Because of the anisotropy of the DDI itself or lattice geometry, which conspire with quantum fluctuations, classical order even at zero temperature is prevented. Such systems, coined quantum spin liquids~\cite{Balents2010,Zou2017,Kurn2018}, exemplify some of the most demanding problems in quantum magnetism~\cite{Diep2012Nov,Schmidt2017}. Adding disorder to the periodic lattice, either by projecting a disordered potential landscape or by relying on random vacancies naturally driving disorder in the system, opens up an opportunity to investigate many-body localization, a phenomenon largely unexplored~\cite{Lukin2014}. Finally, naturally embedded in the DDI is also spin-orbit coupling (SOC) with a novel many-body character~\cite{Syzranov2014}. The excitations generated by such SOC possess non-trivial topology and a Berry phase reminiscent to the one responsible for the unconventional quantum Hall effect in bilayer graphene~\cite{Falko2006}. Uncovering the above phenomena in the laboratory will elucidate their exotic properties, challenging their theoretical description and most probably will also lead to unexpected findings.

Under the ultracold atomic physics umbrella, there exist a few experimental platforms to tackle the above physics. One of the most mature approaches is based on ultacold diatomic molecules, exploiting the existence of nearly-degenerate rotational opposite parity states (OPS) in the ground electronic manifold ($^1\Sigma$) of such linear rigid rotor molecules. The atomic constituents of these diatomic molecules are typically chosen from the alkali family due to the ability to prepare them at degeneracy~\cite{Moses2015Nov,Park2015, Takehoshi2014brief, Molony2016Nov,Seelberg2018,Guo2016}. The OPS can subsequently be polarized with static electric fields or microwave (MW) radiation, which essentially provides a system of electric dipoles oriented in the laboratory frame in the vicinity of few Debye (at full polarization 0.1~D for LiNa to $\sim$4~D for NaCs, LiCs). Additionally, tunability of the dipole strength and even advanced tailoring of the dipolar potential itself~\cite{Micheli2007,Gorshkov2011,Micheli2006atf} is possible. The only drawback is the extreme complexity and the difficulty to prepare samples at low enough entropy and therefore a filling fraction needed to enable percolation~\cite{DietrichStauffer2018}. 
High-lying Rydberg states in atoms provide yet another example of a relatively young and versatile toolbox toward engineering of various spin models~\cite{glaetzle2015,Pohl2015,Whitlock2017,Wu2021}. The short lifetime of laser-accessible Rydberg states due to spontaneous decay and their susceptibility to MW blackbody state transfer have been remedied by utilizing high-lying circular Rydberg states, cryogenic environments and even spontaneous emission inhibiting structures~\cite{Nguyen2018}.    

Many proposals exist also with simple alkali atoms where the magnetic interaction appears as a super-exchange process which scales as $t^2/U$ ($t$ is the tunneling rate, $U$ on-site interaction energy). The fact that the magnetism is generated through motion, demands temperatures in the pK range~\cite{Lewenstein2007}. Atoms loaded in a $p$-band utilizing the same super-exchange mechanism can realize even more demanding models of magnetism, where apart from low temperatures needed, an additional predicament is the unit filling of the $p$-band~\cite{Fernanda2013}.

Atoms with a high magnetic dipole moment offer another path for the same objective~\cite{Chomaz2022}. The simplicity of working with an atom is highly auspicious although there is an added complexity in order to achieve tuning of the dipole strength and change the sign of the DDI~\cite{Lev2018} needed to tune the rates in the particular spin model. The magnetic moment is also weaker than its electric counterpart, reaching an equivalent of 0.1~D for the most magnetic atom of Dy. The above systems are also conceptually different, especially if multiple Zeeman quantum states are allowed to participate in the DDI. Namely the induced electric dipole scaling with the applied field compared to its magnetic counterpart is different, given that the last one is always on. Some unique to the magnetic moment experiments were performed on dipolar relaxation~\cite{Pasquiou2010} and resonant demagnetization in dipolar condensates~\cite{dePaz2013}.
Gases with both magnetic and electric dipoles have also been discussed and novel possibilities for engineering quantum droplets in such doubly dipolar Bose-Einstein condensates were outlined~\cite{Mishra2020,Ghosh2022}.

\begin{figure}
\includegraphics [width=\columnwidth]{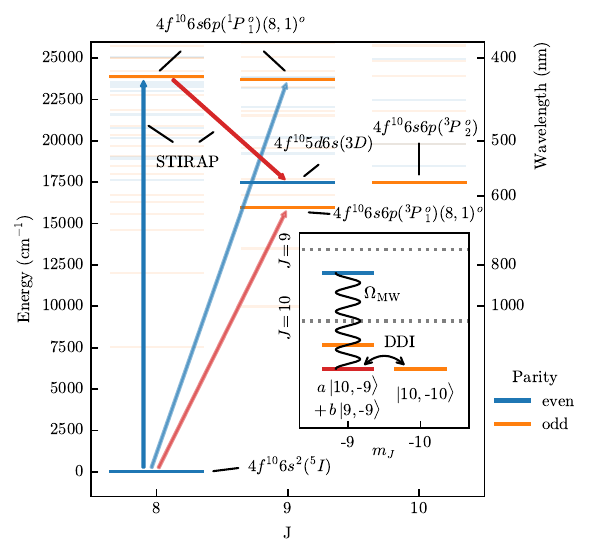}
\caption{\label{levels_updated} Relevant atomic states and optical transitions used in the experiment. Shown are the main cooling transitions at $421$~nm and $626$~nm and one possible STIRAP path to the OPS. The insert shows relevant magnetic sublevels of the OPS, exhibiting a downshift from their original position (dashed line) in a magnetic field. The linearly polarized microwave field (Rabi frequency $\Omega_\mathrm{MW}$) leads to a dressed state (red) with contributions from the $|10,-9\rangle$ and $|9,-9\rangle$ states. The weights $a$ and $b$ depend on the MW Rabi frequency and detuning (details in the Appendix~\ref{isospin detail}). The dressed state and the unperturbed $|10,-10\rangle$ can be tuned to a degenerate isospin-1/2 system, coupled with both eDDI and mDDI.}
\end{figure}

In this article, we describe in detail a new cold-atom experimental apparatus based on Dy. Our main research goal is to simulate spin models of quantum magnetism. To achieve these goals we are set up to pursue a familiar strategy, utilizing the high magnetic moment available in the ground state of the Dy atom. Additionally, we offer an alternative platform, which is attractive for its simplicity and diversity and offers some further means to approach an even more exotic spin system, currently out of reach in the ultracold atoms community or demanding great experimental complexity. The platform is based on a set of closely-spaced OPS in Dy, that can be efficiently populated by a Stimulated Raman Adiabatic Passage (STIRAP) technique, and subsequently manipulated with MW radiation. The ensuing toolbox borrows from both exhaustive magnetic DDI (mDDI) and electric DDI (eDDI), including even the embedded spin-orbit interaction constituent. This last component allows for the simulation of the generalized $XYZ$ spin model, with independently tunable and lattice-dependent rates. 

Experimentally, the apparatus contains a few unique and uncommon design features and engineering choices. A main attribute is a high-resolution CAM, designed for resolving quarter-micron sized features, comparable to the site spacing in short wavelength optical lattices. Another unique feature is the compactness of the apparatus, containing the laser cooling region and the CAM within the same glass vacuum vessel. 

Generally, the bulk of the article describes prior measurements and optimization procedures using the $^{164}$Dy isotope, as a prerequisite for pursuing the above goals.  
The paper is organized as follows: First, as a main motivation, we present two approaches capable of simulating spin models of quantum magnetism and discuss their respective advantages. The second approach based on the OPS has been given the major attention in Sec.~\ref{Survey of experimental possibilities} due to its unique features. We then describe in detail the apparatus itself (Sec.~\ref{Experimental setup}), focusing on its main constituents: vacuum system, magnetic field coils, laser systems and CAM. Section~\ref{Experimental procedure} describes the steps necessary to achieve quantum degeneracy and prepare a 2D sample of Dy atoms in the focal plane of the CAM: transverse cooling and Zeeman slower, magneto-optical trap, evaporative cooling and transport and characterization of optical lattices. In Sec.~\ref{Fluorescence imaging with the CAM} we discuss the first preliminary measurements with the CAM. Finally, in Sec.~\ref{Conclusions and Perspectives} we discuss perspectives and conclude.

\section{Experimental prospects and Motivation}
\label{Survey of experimental possibilities}
We would like to begin by underlining the flexibility of the current setup. We are set to pursue a well established approach to simulate quantum spin models using the ground state magnetic moment of Dy of $10\mu_B$~\cite{Patscheider2020, Chomaz2022}. This approach has both scientific and technical advantages. The long lifetime of the dipoles and the possibility to study, on the same ground, spin-spin interaction and tunneling is an asset of this approach. On the other hand, to generate significant mDDI one may need lattices with spacing in the UV range, for which our apparatus is suited. In the ground state of Dy one can isolate a region in the UV in the range 350-385~nm where the dynamic polarizability (DP) is negative and large (-1000 to -500 a.u.). Furthermore it is free of resonances and heating should be minimal. This approach may bring new challenges, such as possible ionization when the atoms are cycled during the imaging process in such a deep UV lattice, the unknown DP of the excited state of the imaging transition in the UV, and, most importantly, the feasibility to directly resolve UV-lattice spacings with our CAM. The last problem may be remedied utilizing super-resolution techniques~\cite{Rust2006Oct,Huang2008Feb}, an approach pursued by other groups~\cite{Alsolami2012Auth,Fraxanet2022}. Other powerful approaches to circumvent the resolution limitations of the CAM are to use a dynamical expanding lattice~\cite{Su2023, Su2024}, or magnification techniques based on matter wave optics~\cite{Asteria2021}.

A different, more exotic experimental platform, inspired by spectroscopic findings in~\cite{Lepers2018}, can be pursued in parallel with the above scheme based on ground state mDDI. The proposal capitalizes on the exceptional existence of nearly-degenerate OPS in the Dy electronic structure that offer long lifetime and large reduced transition electric and magnetic dipole moments of 8~D and $13\mu_B$ respectively. The states with angular momentum $J=10$ ($[Xe]4f^{10}6s6p$) and $J=9$ ($[Xe]4f^{10}5d6s$) are spaced 35~GHz apart and can be mixed by a MW or DC electric field (Fig.~\ref{levels_updated}). These specific OPS in Dy were considered as a candidate for parity (P)- and time (T)-reversal violation effects caused by the weak interaction~\cite{Roberts2015}. 
The accidental degeneracy of a similar set of OPS in Dy was also used for tests of time variation of the fine structure constant~\cite{Leefer2013} and violations of Lorenz symmetry and Einstein equivalence principle~\cite{Hohensee2013}.

Departing from the use of OPS in Dy mainly as a precision measurement tool, we concentrate on the possibility it provides as a host of an isospin-1/2 basis and its relation to the two-body DDI of lattice pinned atoms. We assume that if two Dy atoms prepared in the OPS end up in the same lattice site a rapid inelastic loss will occur which will cause the atoms to leave the trap. Therefore, for models utilizing the OPS we have to assume a hard-core constraint based on the quantum Zeno effect, meaning no more than one particle per well can reside, similar to chemically reactive molecules~\cite{Yan2013Sep}. 
\begin{figure*}
\includegraphics [height=5.4cm, width=15cm]{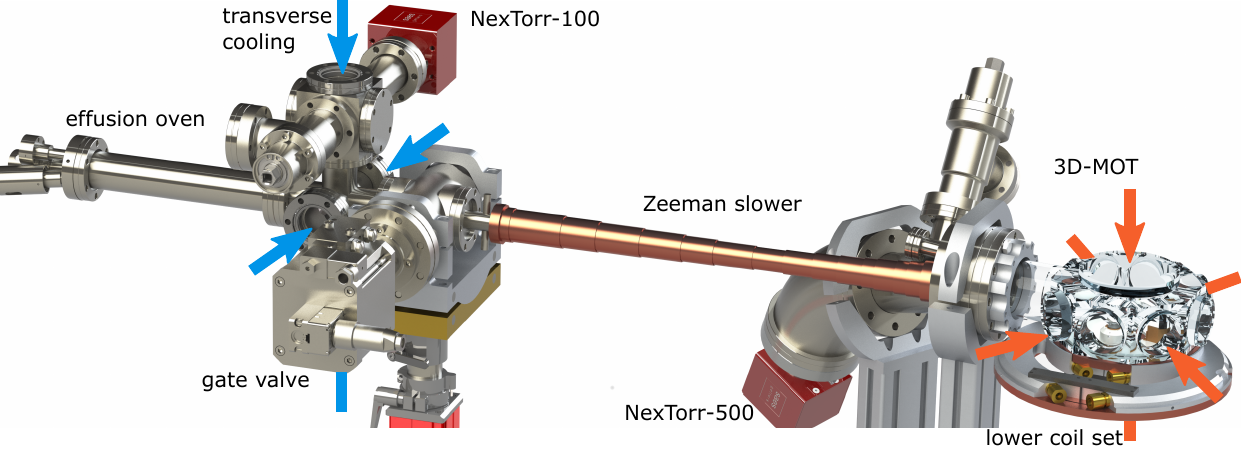}
\caption{\label{ExperimentalSetupFullTransparent} View of the main vacuum apparatus, consisting of a Dy atomic-beam oven, transverse cooling section, ZS and main glass cell. Regarding the magnetic field coils, only the bottom combination under the glass cell (two sets of gradient and offset coils for both MOT and CAM regions) is shown.}
\end{figure*}

We envision a scenario where two MW dressed states composed of magnetic sublevels from the above OPS can be brought to degeneracy and subsequently coupled by both mDDI and eDDI (Fig.~\ref{levels_updated} inset). In our example, we mix the $J=10, m_J=-9$ and $J=9, m_J=-9$ states with a linearly polarized MW, which leads to a dressed state that has contributions from both parities. We then induce a level crossing between the dressed state and the $J=10, m_J=-10$ state by utilizing the linear Zeeman shift at some magnetic field. In the formed degenerate isospin-1/2 basis, additionally to mDDI, eDDI coupling is possible due to the opposite parity component of the dressed state (see for details the Appendix~\ref{isospin detail}), which can be tuned by the MW Rabi frequency and detuning.
One can then write such a long-range interacting system as an extended $XYZ$ spin model, with tunable and lattice dependent rates.

Similar platforms have been theoretically proposed only with complex laser-dressing schemes of Rydberg atoms~\cite{glaetzle2015}, diatomic molecules with $^2\Sigma_{1/2}$ electronic ground state~\cite{Micheli2006atf} or with symmetric top molecules~\cite{Wall2013Nov, Wall2015Feb} (such as methyl fluoride CH$_3$F or hydroxyl radical OH). In the last case, the OPS consists of neighboring opposite parity rotational levels that can be brought to degeneracy via a linear Stark effect~\cite{Wall2015Feb}, provided that the molecule has a non-zero projection of the rotational angular momenta on the symmetry axis. However, the resulting DDI interaction is solely an eDDI. In contrast, the additional mDDI in our system adds another degree of freedom to tune the spin rates of the model. The aforementioned systems still have to be brought to quantum degeneracy in order to realize their full scientific and technological potential~\cite{Vilas2022, Hallas2022}.

In general, the $XYZ$ spin model in 2D is computationally very intensive to simulate for large systems. In the limiting case of 1D the predicted phase diagram is already extremely rich~\cite{Fernanda2013} containing a gapless floating phase non-existent for the symmetric $XXZ$ spin model. The transitions to this phase are also intriguing, including a BKT transition to an antiferromagnetic (AFM) phase on one side and a commensurate-incommensurate transition to a spin-flip phase on the other~\cite{Bak1982}. The width of this phase is proportional to the asymmetry in the $X$ and $Y$ rates~\cite{Fernanda2013}. In 2D, a classical Monte Carlo simulation of few lattice sites of the $XYZ$ model with added Dzyaloshinskii-Moriya interaction reveals a rich interplay between magnetic orders and spin spiral orders~\cite{Gong2015}.

Experimentally, populating the OPS can be performed with STIRAP or Raman pulses. We have chosen a $\Lambda$-level structure incorporating the excited state whose $[Xe]4f^{10}6s6p(8,1)^{\circ}$ character and $J=8$ angular momentum provides a decent transitional dipole moment of 0.3~a.u.\ to the $J=9$ OPS member (Fig.~\ref{levels_updated}). We can draw conclusions about the DP of the $J=10$ state of the OPS from a nearby state with the same character at 15972~cm$^{-1}$ above the ground state~\cite{LepersPR}. At 1064~nm the DP was calculated to be $\sim$140 a.u., which is comfortably close to the measured ground state value of 184~a.u.~\cite{Ravensbergen2018May}. Additionally, the expected large vectorial and tensorial parts of the DP could be exploited, to achieve magic wavelength conditions for the ground state and OPS such as proposed for other Dy  transitions in Ref.~\cite{Dzuba2011}. 
To investigate the above phases we have developed a sensitive diagnostic tool, namely a CAM, which should be capable of single atom spin- and site-resolved $in\: situ$ detection and therefore measurement of multi-point spatial correlations~\cite{Gross2021Dec}. The design optical resolution of $0.23~\mu$m and similar depth of focus of the CAM, makes the apparatus particularly tuned to a single or double layer 2D structures.   

\section{Experimental setup}
\label{Experimental setup}
In this section we give a detailed overview over the main experimental tools. In Sec.~\ref{General experimental structure} we discuss the general anatomy of the experiment. We then describe in detail our vacuum system and magnetic field coils (Sec.~\ref{Vacuum system and magnetic field coils}), laser systems (Sec.~\ref{Laser system}), CAM design (Sec.~\ref{Quantum gas microscope}) and optical lattices implementation (Sec.~\ref{Optical lattices implementation}).

\subsection{General experimental structure}
\label{General experimental structure}
The main apparatus consists of an effusive oven, immediately followed by a transverse cooling (TC) section, Zeeman slower (ZS) and a single glass cell, dedicated to the main cooling and trapping region and containing the CAM itself. There are two laser cooling systems, two STIRAP laser systems, and three high power lasers for dipole traps and generation of the optical lattices. All the lasers that need to be frequency stabilized are referenced to an ultra-low expansion (ULE) cavity. All the necessary electronics consisting of digital and analog cards, direct digital synthesizers (DDS), servo electronics (PIDs) for laser frequency, power and coils current stabilization are home made and controlled by a digital  card (NI 6533 Digital I/O Card). The experiment control software, implemented in Visual C++~\cite{FlorianS}, writes the commands on the NI card (operating at 1~MHz) and communicates with the data acquisition computer. The software for setting the camera properties, acquiring and presenting the atomic images and data analysis, is self-written in Python.
\begin{figure*}
    \includegraphics [height=8.7cm, width=13cm]{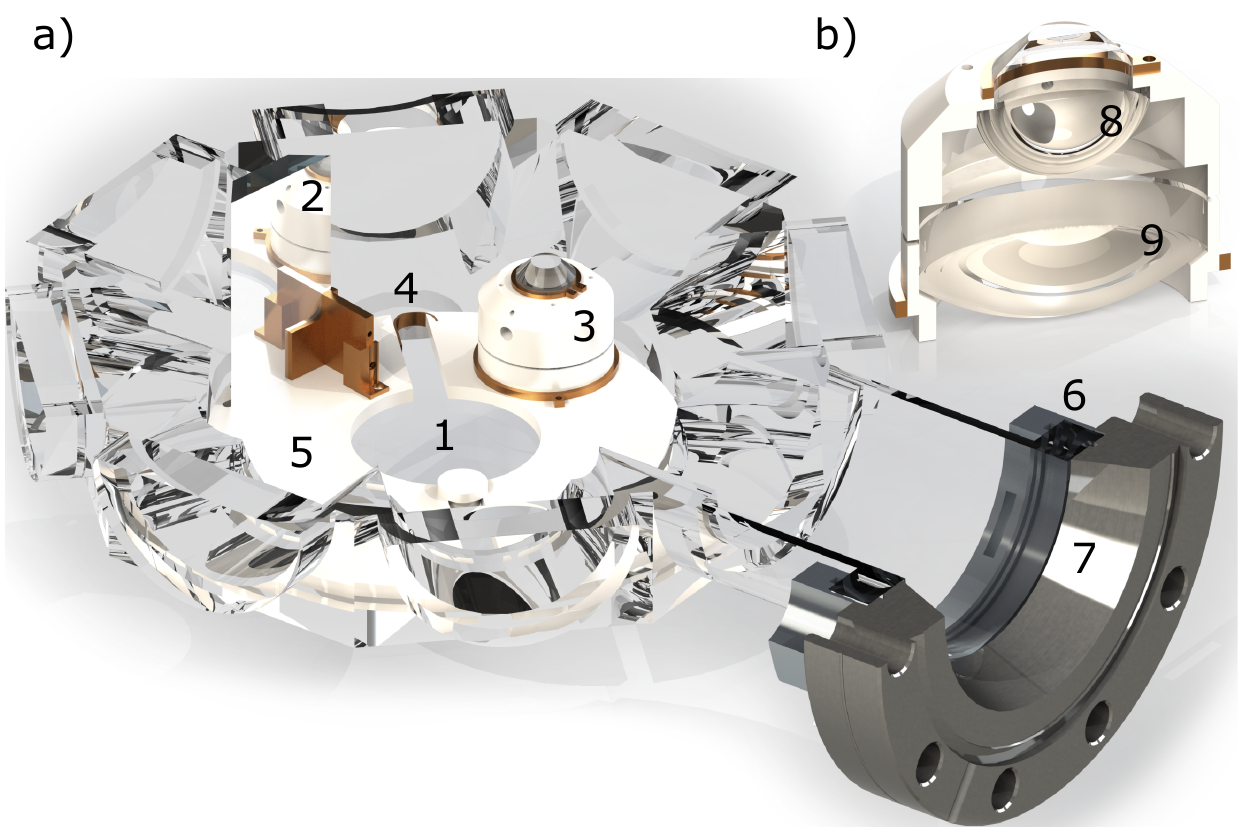}
    \caption{\label{cell_and_CAM} Anatomy of the main glass cell illustrating a) the positioning of laser cooling region (1), ZS mirror (2) and CAM (3). Two spring loaded (4) Macor semi-circles (5), provide a stable base for positioning and securing of components (see text). The connection to the rest of the vacuum chamber is accomplished via an indium seal (6) and a custom adapter flange (7). b) A cross section of the CAM, consisting of a beveled front plano-spherical lens (8) and a second plano-aspheric lens (9). The two lenses are held with a Macor enclosure.}
\end{figure*}

Three cameras (labeled 1 to 3) are used to take images at different Dy cloud locations in the glass cell. In the central region of the cell, where laser cooling and trapping takes place initially, we use absorption imaging both horizontally and vertically with camera 1 (Andor 5.5 Zyla sCMOS). To precisely position the sample above the CAM we perform both absorption and fluorescence imaging horizontally with camera 2, identical to camera 1. The CAM is oriented vertically (see Fig.~\ref{cell_and_CAM}), and collects the fluorescence signal from the atoms and images it onto camera 3 (Andor-iXon-897-EXF).

\subsection{Vacuum system and magnetic field coils}
\label{Vacuum system and magnetic field coils}
The vacuum system is separated in three sections (Fig.~\ref{ExperimentalSetupFullTransparent}). The first section contains an oven (WEZ 40-10-22-K-HL-C-0, MBE Komponenten) used generally for molecular beam epitaxy. It has a motorized shutter, an independent hot lip temperature control  and a water-cooling jacket positioned around a tantalum crucible, containing pellets of Dy (about 15~g). The front of the crucible (length 100~mm, $\diameter$12~mm) has a long thin tube (length 40~mm, $\diameter$2.5~mm) to collimate the Dy beam. The shutter is opened only during the magneto-optical trap (MOT) loading stage. At this ratio of the length to diameter of the tube one theoretically expects a $\pm$4.2$^{\circ}$ atomic cone. The oven is followed by a TC section formed by a standard vacuum cross, with four anti-reflection coated (AR) glass ports (Vaqtec). In this section we use a NexTorr D100-5 combined pump with getter (St-172, 100~l/s for H$_2$) and embedded ion pump (6 l/s for Ar). Given that the Dy deposited on the walls also serves as an additional getter the pressure in this section is less than 1$\times 10^{-11}$~mbar, as indicated by the vanishing ion pump current. We don't incorporate additional ion gauges to monitor the pressure to decrease complexity and cost. Before entering the ZS section, the Dy beam crosses the orifice of an all-metal pneumatic gate valve (VAT GmbH, 48132-CE44-0002). The valve provides necessary detachment of the oven from the main apparatus during a Dy reloading procedure, and allows for an emergency shut down if the pressure in that region exceeds an alarming value. 

The next vacuum section is the ZS, which consists of two stainless steel (alloy 316L) cylindrical pieces with an internal conical surface, formed by electrical discharge machining. The pieces are welded together in one piece with total length of 560~mm with small and large diameters of 4 and 14~mm, respectively. This conical shape matches, but doesn't restrict, the atomic beam shape leaving the exit aperture of the effusive oven and arriving at the trapping MOT position. On the outside, the volume remaining between the conical surface and an enclosing cylinder serves as a water cooling section. A simple geometric constraint forces the water flow to travel along the length of the cylinder twice before exiting, and therefore assures proper cooling along the full length.

The third vacuum section is a fused-silica glass cell (Precision Glassblowing, 8 horizontal ports $\diameter$50.8~mm, top and bottom ports $\diameter$101.6~mm, see Fig.~\ref{cell_and_CAM}a) in the shape of an octagon, with one side port used as entrance for the atoms and the rest for laser cooling, dipole traps and optical lattices. All windows are corrugated with surface relief microstructures providing AR coating with reflection coefficient $<0.25\%$ in the range 250-1600~nm and at angles of incidence 0-60$^{\circ}$ (produced by Telaztec). The standard glass to metal transition used in the vacuum industry to bridge between glass vessels and a ConFlat (CF) vacuum seal standard, for this size cell, would be impractically long. Therefore, we designed a custom adapter flange, which on one side contains the CF seal and connects to the glass on the other side with an indium seal. The indium seal is made out of round cold-bonded indium wire ($\diameter 1$~mm) pressed between the adapter flange and a flat glass lip at the entry of the cell (Fig.~\ref{cell_and_CAM}a). With the current design the distance from the end of the ZS to the MOT center is 23~cm, substantially longer compared to standard setups~\cite{Muhlbauer2018May, Izhofer2018}. The vacuum here is provided by a getter-ion pump (NexTorr D500-5). The pump is positioned about 25~cm from the glass cell at a 45$^\circ$ angle to the cell entrance, in order to avoid stressing the cell thermally during the getter activation at 600$^{\circ}$C. Nevertheless, due to the relatively large tube (CF63) used, the pumping speed for hydrogen at the MOT center is almost completely utilized. The pressure in the cell is also
$<1\times 10^{-11}$~mbar, shown by the current of the ion pump but also by the lifetime of a Dy Bose-Einstein condensate (BEC). 
During the initial bake-out the cell was heated to $120^{\circ}$C for two weeks by wrapping a heat tape around an aluminum enclosure concentric with it. A simple thermal simulation assured us that the cell does not experience large thermal gradients during the NexTorr activation. 

The glass cell, made out of pure glass with no bolting structures, is not suited for positioning in-vacuum components. We solve this problem by positioning two semi-circles out of machinable ceramic (macor, Corning Inc.) with devoted cuts to accommodate the MOT light and the positioning of the desired in-vacuum elements (Figs.~\ref{cell_and_CAM} and~\ref{transport_cross_section}). The two semi-circles were carefully placed in the cell with a vacuum suction tip attached to a 3D precision linear micrometer stage. They are spring-loaded and once put down on the bottom glass surface they seize to the side surfaces. The adhesion to the bottom glass and seizing to the side surfaces creates a stable base and allows positioning of the CAM and a ZS in-vacuum mirror into devoted slots. The ZS mirror is custom-made with pure aluminum deposited on a fused silica substrate, without protective layer (Figs.~\ref{cell_and_CAM} and~\ref{transport_cross_section}).

In order to compensate ambient magnetic fields and to apply small ($<2$~G) uniform offsets, three pairs of large compensation coils are symmetrically installed around the glass cell. The coils have the shape of rectangular cuboid ($xyz$ dimensions $944\times1174\times674$~mm$^3$, coordinates defined on Fig.~\ref{transport_cross_section} as a reference for the whole paper), and are driven by a home-made bi-polar current source based on high-current paralleled operational amplifiers. The fields obtainable with these coils are $(0.8,~1.5,~2.15)$~G for the $xyz$ directions. We additionally employ two sets of offset/gradient coils (Fig.~\ref{ExperimentalSetupFullTransparent}), concentric with the $z$-axis of symmetry of the MOT and CAM regions respectively. The offset coils for both the MOT and CAM regions are in Helmholtz configuration. The water-cooled holding structure resembles two circular loops (with centers 34~mm apart, to match the distance between CAM and MOT axis), touching internally at a point. Additionally, we have an option to apply a uniform magnetic field in the $y$-direction (Fig.~\ref{transport_cross_section}) with an independent set of coils centered around the CAM $y$-symmetry axis. The obtainable fields and field gradients of those coils, driven by identical bi-polar supplies as for the compensation coils, are $(B^z_{0},dB^z_{gr}/dz)$=(18~G, 3~G/cm) and $B^y_{0}$=6~G. To prevent induced eddy currents in all metallic structures forming continuous loops nearby the experiment, such as aluminum structures holding the coils, they are sliced and the removed material is substituted with a non-conductive plastic joint to maintain the structural stability. 
\subsection{Laser systems}
\label{Laser system}
\begin{figure}
\includegraphics [height=7cm, width=8cm]{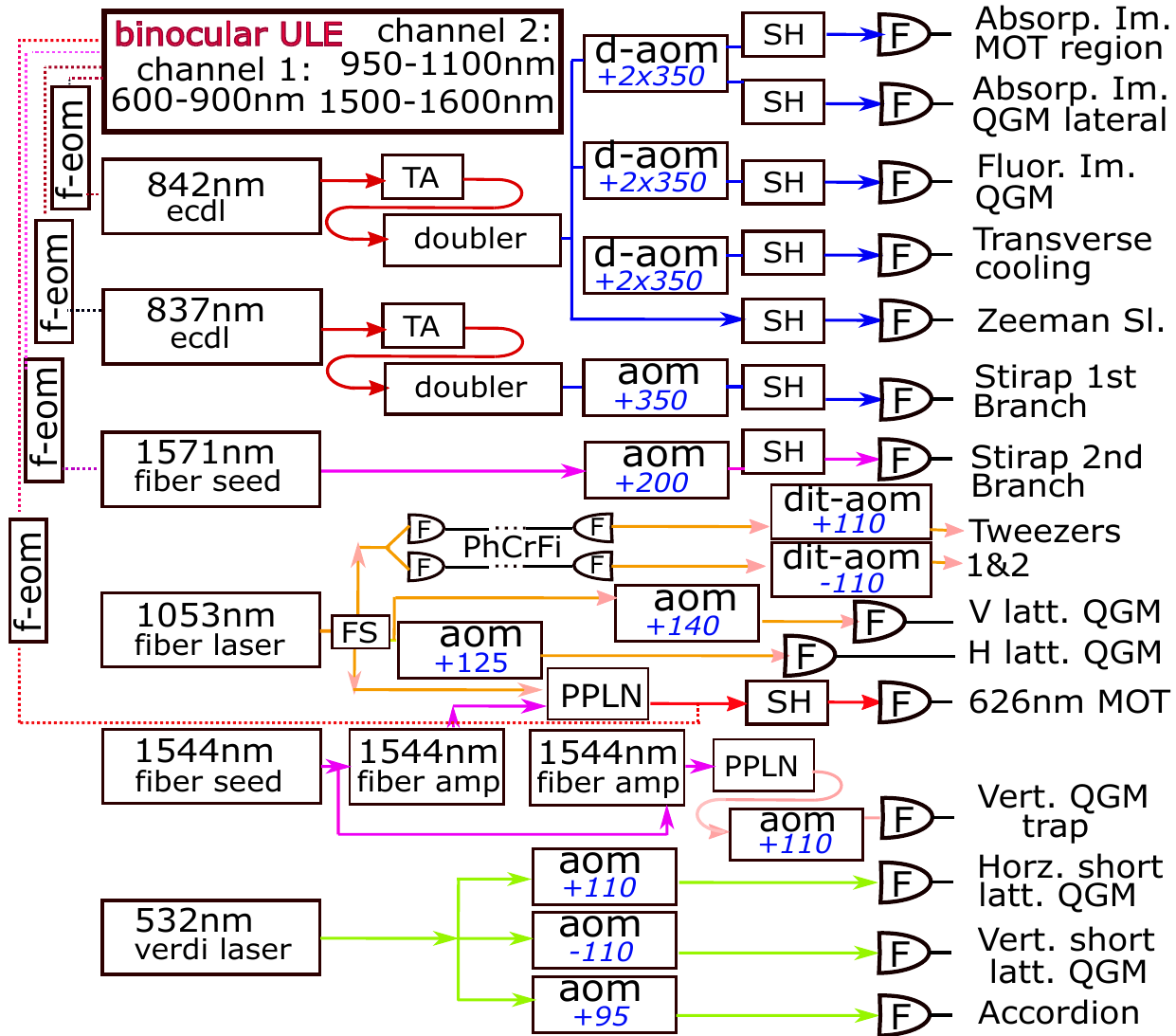}
\caption{\label{laser_systems} Diagram of the laser systems used in the experiment. Abbreviations not defined in the main text: FS- motorized fiber splitter, F- fiber launch, PhCrFi- photonic crystal fiber, d-aom - double pass AOM, SH- mechanical shutter, dit-aom- dithering AOM }
\end{figure}

The laser systems used in the experiment are depicted in Fig.~\ref{laser_systems}. The relevant transitions for laser cooling are also shown in Fig.~\ref{levels_updated}. We separate the laser systems in three different groups based on the technology employed. 

The first group is based on diode laser technology. We use two home-made, amplified and frequency-doubled diode laser systems for ZS, TC, imaging and for the first STIRAP branch. We operate ZS, TC and imaging on the 421-nm transition $4f^{10}6s^2(J=8) \leftrightarrow 4f^{10}6s6p (J=9)$, which has a natural linewidth $\Gamma_{421}$=2$\pi\times$32.2~MHz~\cite{Lu2011} and saturation intensity $I_{421}=56.3$~mW/cm$^2$. The laser system consists of a cat-eye extended cavity diode laser (ECDL,~\cite{Thompson2012}) operating at 842~nm (Eagleyard AR coated diode, 50~mW), tapered amplifier (TA, Eagleyard, 2-W chip) and a bow-tie frequency doubling stage based on an LBO crystal (Raicol Crystals). The ECDL delivers 35~mW to the TA which amplifies it to 1.4~W of useful output before the frequency doubler. We utilize type-I critical phase-matching and on resonance attain 600~mW of blue light at 421~nm. We observe a slight deterioration of doubling efficiency over the course of a year, possibly related to the coating of the crystal, although the doubler is in a sealed box. The doubling cavity is stabilized relative to the ECDL with a polarization-based H{\"a}nsch–Couillaud technique~\cite{Hansch1980}. The laser system for the first STIRAP branch addresses the $4f^{10}6s^2(J=8)\leftrightarrow 4f^{10}6s6p (J=8)$ transition at 418.8~nm~\cite{Lepers2018} and is identical to the one described above, except because of the low power requirement, a low power 1-W TA chip (Eagleyard) is employed. We obtain 40~mW at 418.8~nm after the second harmonic generation.  Both of the ECDLs of the above systems are frequency locked using a Pound-Drever-Hall (PDH,~\cite{Drever1983}) method to a home-built optical resonator.
\begin{figure}
\includegraphics [height=11cm, width=8cm]{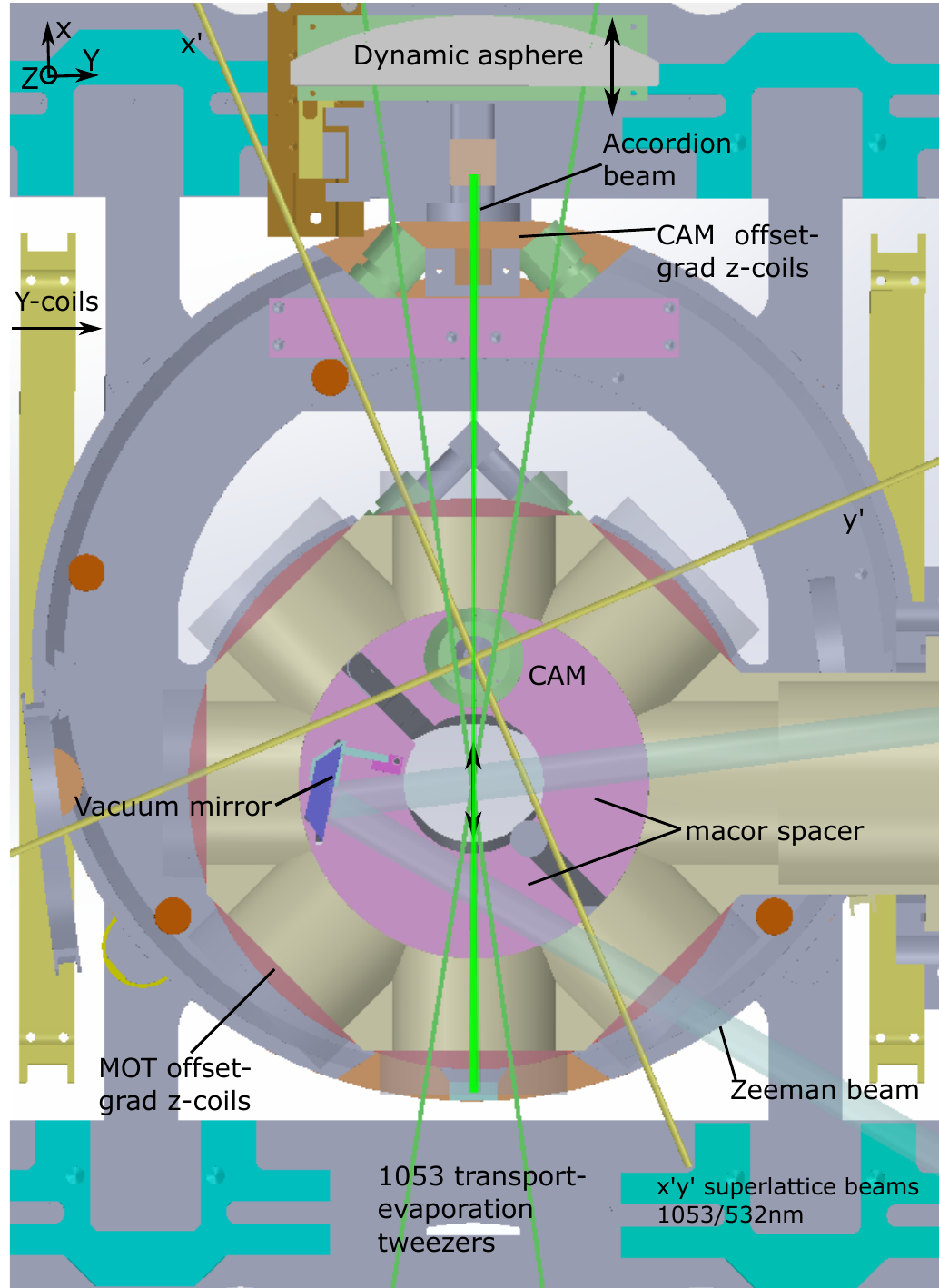}
\caption{\label{transport_cross_section} Cross section of the main cell illustrating the two spring loaded macor platforms forming a base for the ZS mirror and CAM. The main tweezer beams performing the trapping, evaporation and transport are shown. Their crossing point moves from the MOT region (center of the cell) to the point on the symmetry $z$ azis of the CAM and 2.6~mm above its focal plane. Lattice beams both at 1064~nm (currently 1053~nm) and 532~nm crossing above the CAM are labeled as $X'Y'$ beams. The two green interfering 532-nm beams, propagate in the $x-z$ plane and cross above the CAM, forming a long wavelength lattice (called FA, see text) of 10~$\mu$m.}
\end{figure}

The resonator (Fig.~\ref{laser_systems}, binocular ULE), which serves as a reference for all stabilized lasers, is based on a binocular ULE glass spacer (Hellma, 1.5~GHz free spectral range) positioned in a vacuum vessel ($10^{-9}$~mbar), and optically contacted partial reflectors (Lens Optics, superpolished) which are coated (LaserOptik GmbH) in the bands of interest to a finesse of $\sim 30,000$. The detuning relative to the ULE, and the sidebands necessary for the PDH method, are generated by a single fibered electro-optic modulator (f-EOM) preceded by a non-linear transmission line (NLTL). The NLTL creates a sawtooth wave from a single frequency, which the f-EOM afterwards converts to a single sideband, bridging the gap between the atomic line and the ULE resonance. The PDH single frequency (10~MHz) signal is added to the NLTL sawtooth wave by a broadband mixer. Eventually a highly tunable (300-1700~MHz) single sideband, attributed with a PDH modulation is achieved, allowing for convenient tuning over the full free spectral range of the ULE cavity~\cite{Kirilov2015}. 

The second group of laser systems are commercial fiber lasers used as dipole traps, optical lattices,  narrow-line MOT light and the second STIRAP branch. The central laser system is a 50-W, 1053-nm fiber laser (IPG Photonics). Most of the power (90$\%$) is devoted to a pair of optical tweezers used for trapping, evaporation and transport of the atomic cloud (detailed in Section~\ref{Experimental procedure}) and additionally for 3D optical lattice beams (Fig.~\ref{transport_cross_section}). The remaining 5~W are frequency summed with a 5-W, 1544-nm single frequency laser (Keopsys, 10~W) to obtain 626-nm light for the narrow line MOT (natural linewidth $\Gamma_{626}=2\pi\times136$~kHz~\cite{Gustavsson1979}). The 1053-nm and 1544-nm amplifiers are both seeded by 5-mW single-frequency fiber lasers (Koheras Adjustic by NKT Photonics). We use a periodically poled lithium niobate crystal (PPLN, Covesion), phase matched at 177.4$^{\circ}$C in a single-pass configuration and derive 1.5~W of 626-nm light, which is then distributed to the vertical and horizontal MOT beams. The 626-nm light is frequency stabilized to our ULE cavity with PDH technique and the correction signal is fed to the fast piezo actuator regulating the cavity of the 1544-nm fiber seeder. 
The 1544-nm seeder additionally powers another fiber amplifier at 10~W (Keopsys) which is frequency doubled to 772~nm in PPLN (Covesion) at 141.8$^{\circ}$C. We obtain 0.7~W useful power for a vertical dipole trap for the atoms at the CAM region (the CAM is AR coated in the range 700-900~nm), by propagating along its symmetry axis. Additionally, we developed a time-incoherent laser system centered at 770~nm (bandwidth FWHM $\Delta\lambda_{770}=9$~nm) based on a super-luminescent diode (Exalos, EXS210017-01), followed by a booster optical amplifier (Thorlabs, BOA780P) and a TA (Eagleyard, EYP-TPA-0765-01500-3006)~\cite{Aaron2021}. The final bandwidth and center are a convolution of the bandwidth of the diode and gain profiles of the booster amplifier and TA. The coherence length is far below the distance of the atoms to the CAM surfaces therefore eliminating any parasitic interference effects from non-perfect AR coating. The system delivers 500~mW post-fiber laser power. 
The last fiber laser is a 40-mW laser at 1572~nm (Koheras adjustic, NKT Photonics), needed for the second STIRAP branch. The laser is inherently narrow ($<1$~kHz) and is simply stabilized with a low-bandwidth servo loop to the ULE cavity using the PDH technique. 

The last laser group consists of a single laser system of a solid-state type, based on a 18-W laser at 532~nm (Coherent Verdi), which is needed for our horizontal short-spacing optical lattice ($x^{\prime}$-$y^{\prime}$ on Fig.~\ref{transport_cross_section}). It also powers a vertical large-spacing lattice of 10$~\mu$m (see Sec.~\ref{Optical lattices}), obtained by interfering two free traveling beams at $\pm$1.5$^{\circ}$ with respect to the horizontal plane. The interference pattern at the crossing point is coincident with the CAM symmetry axis and additionally overlapped with the CAM focal plane. 

\subsection{Cold atom microscope}
\label{Quantum gas microscope}
Our CAM is an in-vacuum compact design made only out of two lenses, a Weierstrass sphere serving as a front plano-covex lens~\cite{Puig2019} and a plano-aspheric lens (Fig.~\ref{cell_and_CAM}b). The design was inspired by~\cite{Robens2017}, and adapted to the blue region, specifically for the blue line at 421~nm of Dy. The Weierstrass sphere enhances the numerical aperture (NA) by a factor of $n^2$ (where $n$ is the refractive index of the sphere), therefore reducing the need for an extreme aspheric surface of the second lens. As a result, the aspheric surface could be manufactured with magneto-rheological finish~\cite{Kumar2020}, achieving the lowest possible irregularity. The optical performance and tolerances of the CAM were simulated and optimized with an optical design software (Zemax 2013). The two lenses were produced by Sill Optics, with both curved surfaces having a PV maximum irregularity of 0.15$~\mu$m and iRMS $<0.05~\mu$m. The enclosure of the CAM is made out of two cylindrical concentric macor pieces. Each one has a very fine, 0.15~mm pitch tap in order to regulate the distance between the two lenses. This degree of freedom was necessary given the lens thickness tolerances and especially the uncertainty of the complex front lens coating (30 interchanging layers of Ta$_2$O$_5$ and SiO$_2$, approximated in the simulation by the total thickness of $2.2~\mu$m of Ta$_2$O$_5$ and $2.9~\mu$m SiO$_2$). In the simulation, the achieved parameters are an NA of $0.92$, working distance of 160$~\mu$m (longer working distances makes the CAM unfitting for our single cell design), field of view $\pm 40~\mu$m and a depth of focus $\pm 125~$nm. The CAM was tested in air by shifting the test wavelength by $\sim 1$~nm to shorter wavelengths, where the performance and distances are the same as for the atomic Dy 421-nm transition when the CAM is positioned in vacuum. We used shear interferometry~\cite{Hariharan2006} on the re-collimated light that crosses the CAM twice after a retro-reflection from a silver mirror positioned in the CAM focal plane. We observed parallel interference fringes with no significant deformation until an aperture of $\diameter=$ 23~mm, corresponding to the desired NA. The light was then focused on a beam-profiler (Thorlabs, BC106N-VIS) with a single lens providing a magnification of 160. The Strehl ratio was deduced across the full field of view, confirming a diffraction limited performance up to the designed NA.  

\subsection{Optical transportation setup}
To be able to reach quantum degeneracy and transport the atoms into the focal plane of the CAM we developed an optical setup capable of moving in 3D a pair of two crossed high power optical dipole traps (we refer to them as tweezers) at 1053~nm, crossing at 14$^\circ$ (Fig.~\ref{transport_cross_section}). The setup needs to be able to efficiently load the Dy cloud from the compressed MOT (CMOT), transport it while performing evaporative cooling, and eventually position it onto the focal plane and symmetry axis of the CAM. 

We utilize a large dynamic aspheric lens, which is a 20~mm wide slice, cut symmetrically around the diameter from a large commercial asphere (Thorlabs AL100200-C, $f=200$~mm, Fig.~\ref{transport_cross_section}). This aspheric lens is positioned in an aluminum frame on an air-bearing stage (ABS) (Aeroflex, ABL1000-050-E2-CMS1-PL2-TAS), capable of transporting the two-tweezer intersection point, over a maximum distance of 50~mm. The intersection point coincides with the asphere focal point during the full transport, due to the parallelism of the tweezers to the lens symmetry axis, before refracting from it. The customized lens cut removes a lot of unnecessary weight and size from the asphere, but still we need carefully positioned counter-weights ensuring that the center of mass of the full structure is within the stage requirements. Such an arrangement allows us to precisely transport the atoms in the focus of the tweezer pair across the 35~mm distance from MOT to CAM regions along the $x$-direction and additionally along $y$ and $z$ (coordinate system on Fig.~\ref{transport_cross_section}). The ABS is slightly magnetic and therefore the holding structure is designed to keep it far from the atoms. 

The  tweezer light is transported onto the experimental table with two photonic crystal fibers (E324-443-500, NKT Photonics), which are water-cooled at the entrance. 
Their power is regulated with two AOMs after the fiber exit, because we actively steer them horizontally at 100~kHz to dynamically broaden the beams at the focus position. 
For this we send an arccosine-shaped signal from an arbitrary waveform generator to a fast voltage-controlled oscillator (VCO, AA Optoelectronic, 1$~\mu$s full range scanning time). 

To set the desired aspect ratio of the tweezers beams, they cross two expanding telescopes made out of a set of cylindrical lenses in order to increase the beam size only along the $z$-direction. The ambient aspect ratio in the absence of the beam steering is 2.5 ($\omega_z=19~\mu$m and $\omega_y=50~\mu$m). The two telescopes are positioned right before the dynamic asphere where the tweezers are parallel to each other and 50~mm apart. The front lenses in each telescope is attached to a miniature precision linear stage equipped with an optical linear encoder (Q-521.130, PI). This allows us to move the tweezers vertically from their original position during the MOT loading stage, at 2.6~mm above the CAM front surface, and lower them to its focal plane at 160~$\mu$m above.
The final necessary degree of freedom along the $y$-direction is not automated and therefore optimized once. The ABS rests on a precisely polished smooth granite block, which is channeled to move along the $y$-direction with two pairs of ceramic ball bearings pressing on its $y-z$ opposite walls. The motion along this final degree of freedom is accomplished by moving the granite block with two micrometers, pressing against its two opposite $x-z$ walls.

\subsection{Optical lattices implementation}
\label{Optical lattices implementation}
Eventually the atoms have to be loaded in a 2D plane coincident with the focal plane of the CAM and pinned by the lattice geometry of choice in that plane. This is both for technical reasons, given the tight depth of focus of the CAM, and fundamental, given the scientific goals of the experiment. As a first step toward preparing a 2D sample we load the cloud into a single plane of a vertical lattice with spacing of 10$~\mu$m, which we refer to as a fixed accordion (FA). The FA is formed by the interference of two beams at 532~nm derived from a 18-W Verdi laser (Fig.~\ref*{laser_systems}), propagating in the $x-z$ plane and intersecting above the CAM at an angle of 3$^\circ$ (Fig.~\ref{transport_cross_section}). The two beams are derived from a single beam incident on a custom coated flat glass substrate positioned at 45$^\circ$ relative to the $z$-axis, with a partially reflective (38$\%$) polarization-independent coating on the front surface and a high reflective coating on the back surface. The high parallelism and choice of reflectivity of the substrate ensures two identical equal-powered and highly parallel beams exiting horizontally (along -$x$) and 5~mm apart (defined by the substrate thickness of 6.35~mm). The two beams are derived from the first reflection and the first transmission of the original beam. About 25$\%$ of the incoming power is lost in the beam remaining after the second exit from the substrate. The two beams are then focused onto the focal plane of the CAM with a $f=100$~mm aspheric lens (Asphericon, AFL25-100-U-U-355-L) and form the interference pattern along the $z$-direction. The interference pattern has a Gaussian envelope with $\diameter 150~\mu$m along $y$-direction and $\diameter 80~\mu$m along the interference pattern in the $z$-direction. The substrate can be moved vertically with sub-micron precision with a slip-stick actuator (PI, N-470 PiezoMike) to adjust the middle fringe of the interference pattern onto the CAM focal plane. 

Currently the laser is not frequency stabilized and the substrate is not actively temperature stabilized, nevertheless the interference pattern is sufficiently stable for the initial characterization of the CAM of this work. Stability and positioning below the depth of focus of the CAM will only be achievable once the atomic 2D plane is referenced to the CAM itself (detailed in Section~\ref{Conclusions and Perspectives}). 
Furthermore for ensuring that the $x-y$ trap frequencies are not severely elevated when increasing the FA power, we have an alternative design based on an interferometer~\cite{Ville2017}, where the horizontal diameter is increased to $\sim 1$~mm. Here, the vertical adjustment of the interference fringes is achieved by a piezo crystal, tuning the length of one of the interferometer branches. To improve the chances of this method for both designs, we image the interference pattern on a 1D charge-coupled device (CCD) camera, and extract an error signal of the drift in position of the central interference fringe. In future we will feed this signal back to an actuator that corrects the FA position. This improvement still does not guarantee a long term stability, and does not cancel vibrations of the interference fringe relative to the CAM. 

To further avoid vibrations and long term drifts, it is instrumental to eventually lock the atomic 2D cloud to the CAM focal plane. 
For this purpose we reflect a 1053-nm vertical lattice beam from the front face of the CAM ($R>99.5\%$ in the range of $1030-1120$~nm from $0-5^{\circ}$ angle of incidence). We discuss in Section~\ref{Conclusions and Perspectives} our future plans to populate a single plane of such a lattice. 

The horizontal $x^{\prime}$$y^{\prime}$ lattices are rotated at 23$^{\circ}$ relative to the $x-y$ coordinate system, due to the available space, and consist of two (for both $x^{\prime}$ and $y^{\prime}$ directions) super-lattice beams of co-propagating 532-nm and 1053-nm beams (Fig.~\ref{transport_cross_section}). 
The retro-reflective optics for each of the super-lattice beams consist of two polarizing beamsplitters with a motorized half-waveplate in between, followed by a mirror. Each of the polarization elements, the mirrors and lenses for shaping the lattice beams are chosen to perform equivalently at both 532-nm and 1053-nm wavelengths. 
With this arrangement, by rotating the waveplate the reflection can be extinguished and the lattice beams become regular dipole traps. This construction simplifies traps alignment procedures and is also used optionally during the experimental cycle to utilize the lattice beams as traps during the evaporation process and later converting them to lattices (the minimum time for the switch, limited by our servo motors, is 100~ms). Each lattice beam is delivered by a photonic crystal fiber identical to the one for the tweezer light, with the difference that in order to avoid reflections back into the fiber the horizontal $x^{\prime}-y^{\prime}$ beams are followed by high-power optical isolators. For the vertical lattice we can avoid the optical isolator, by slightly deviating the incoming beam from the normal incidence and then blocking the back reflection, given the proximity of the atoms to the retro-reflective front surface of the CAM.  

\section{Experimental procedures and system characterization}
\label{Experimental procedure}
In this section we describe a typical experimental procedure, starting with the inital slowing of the atomic beam (Sec.~\ref{Transverse cooling and Zeeman slower}), followed by the MOT stage (Sec.~\ref{Magneto-optical trap}) and loading of the dipole traps with subsequent evaporation and transport to the CAM (Sec.~\ref{Evaporative cooling and transport}). Finally, we characterize the optical lattices in Sec.~\ref{Optical lattices}.
\subsection{Transverse cooling and Zeeman slower}
\label{Transverse cooling and Zeeman slower}
\begin{figure}
\includegraphics [height=8cm, width=8cm]{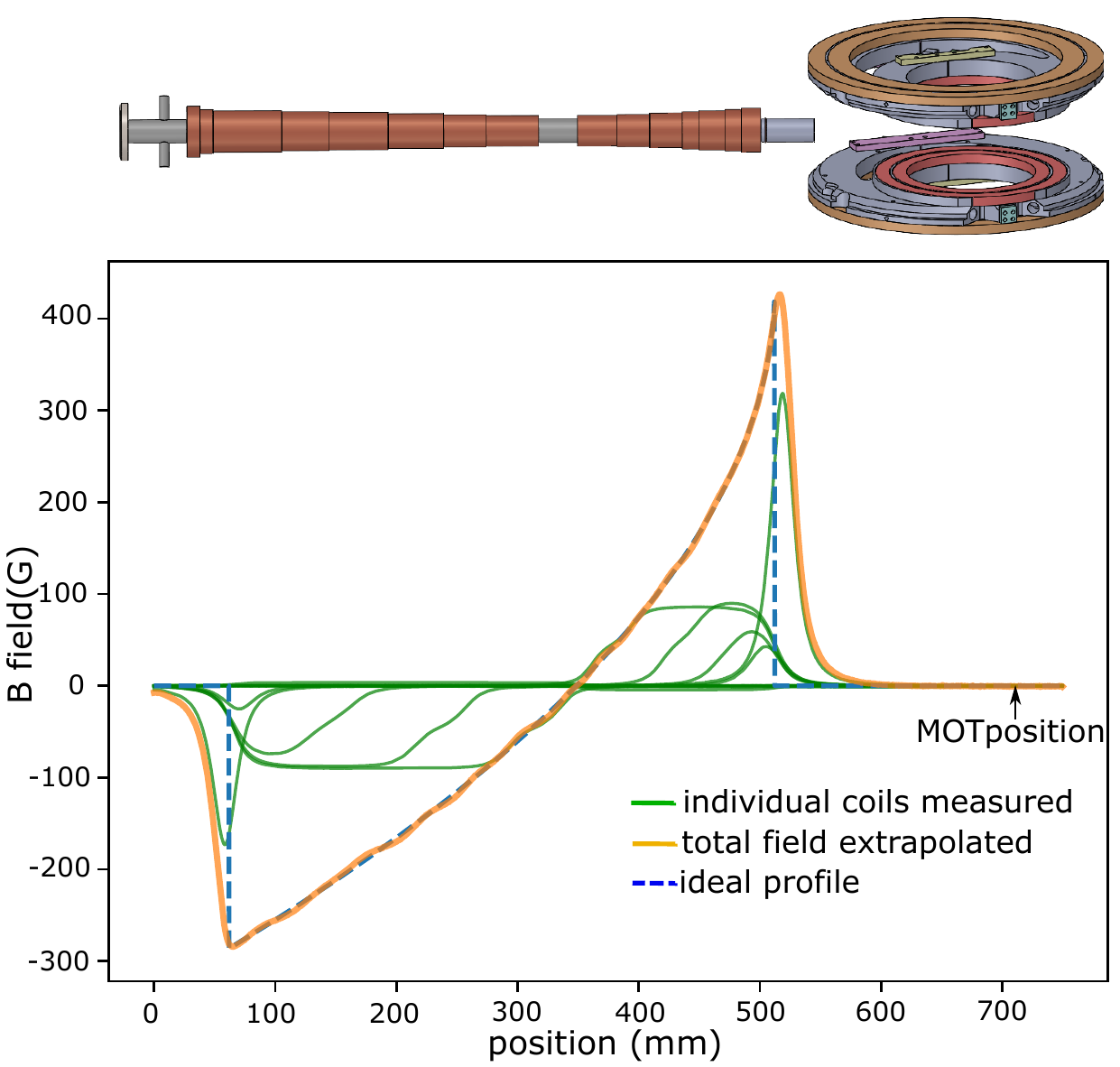}
\caption{\label{zeeman_currents} ZS coils and main set of coils. Each set consists of gradient and offset coils in Helmholtz configuration, concentric with the MOT and CAM $z$-axis of symmetry, respectively but sharing a common water-cooled sink. Presented are also the measured individual fields of each ZS coil, total on-axis field and the desired theoretical field (dashed line).}
\end{figure}

We operate the Dy oven with the crucible and the hot lip temperature at 1000$^\circ$C and 1100$^\circ$C, respectively, to produce a beam with a most probable longitudinal velocity of about 450~m/s~\cite{Youn2010,Frisch2012,Maier2014,Muhlbauer2018May}. Immediately following the oven, before entering the Zeeman slower (Fig.~\ref{zeeman_currents}), we perform TC (via the four glass ports following the oven in Fig.~\ref{ExperimentalSetupFullTransparent}) to decrease the angular spread of the beam~\cite{Leefer2010}. We apply cooling in 2D, by utilizing a orthogonal retro-reflected beam configuration with 30~mW in each direction (elliptical shape with $\diameter 26\times$5~mm$^2$ with long axis along the atomic beam). We experimentally optimize the polarization of the individual beams and find the optimum detuning to be $0.5\Gamma_{421}$, as expected for low saturation. The increase in atom number in the 3D MOT due to the TC is shown on Fig.~\ref{MOTs}a. It is clear that the atom number does not plateau out with laser power, given the available laser intensity. We achieve a factor of 7 increase in atom number in the MOT (described below) at the maximum available power.  

Next the atoms enter the ZS region. Here the atoms experience a resonant 421-nm slowing beam, focused on the oven exit aperture in order to match the diverging atomic beam. We operate at large detuning of $\sim 20\Gamma_{421}$ and send about 150~mW of 421-nm light, which is reflected from the in-vacuum aluminum coated mirror. We observe a vague spot on the mirror due to the deposited Dy and assume that about 20$\%$ of the reflected light is lost, based on experience of other laboratories utilizing a in-vacuum mirror for Dy ZS. The ZS is designed with a safety factor of $\eta=a_{zs}/a_{max}\approx 0.4$ (here $a_{zs}$ is the true acceleration and $a_{max}=5.8\times 10^{5}$~m/s$^2$ is the max acceleration possible). The length is chosen to achieve a targeted final velocity of $v_{f}=7$~m/s to match the MOT capture velocity. Given the large distance from the end of the ZS to the MOT center, we may in future adapt a better strategy, allowing a higher final velocity in order to reduce the transverse expansion and then slow the atoms to the MOT trapping velocity right before the MOT~\cite{Lunden2020}. We first tested the ZS performance before assembly, by mapping the magnetic field of each individual coil ($B_i(z)$, $i$ is the coil index and $z$ is the longitudinal coordinate) first for a current value of 1~A. Subsequently a minimization procedure over this basis is used to find the optimal individual currents $I_i$, realizing a total experimentally optimized magnetic field $B_{ZS}(z)=\sum_iI_iB_i(z)$, matching the ideal theoretical field of $B_{th}=B_b+B_0\sqrt{1-z/z_0}$. Here $B_b$ and $B_0$ are the bias and overall height of the field profile and $z_0$ is the ZS length (Fig.~\ref{zeeman_currents}). Further fine tuning of the currents was done experimentally by directly maximizing the number of atoms in the MOT.

\subsection{Magneto-optical trap}
\label{Magneto-optical trap}
\begin{figure}
\includegraphics [height=15.4cm, width=8cm]{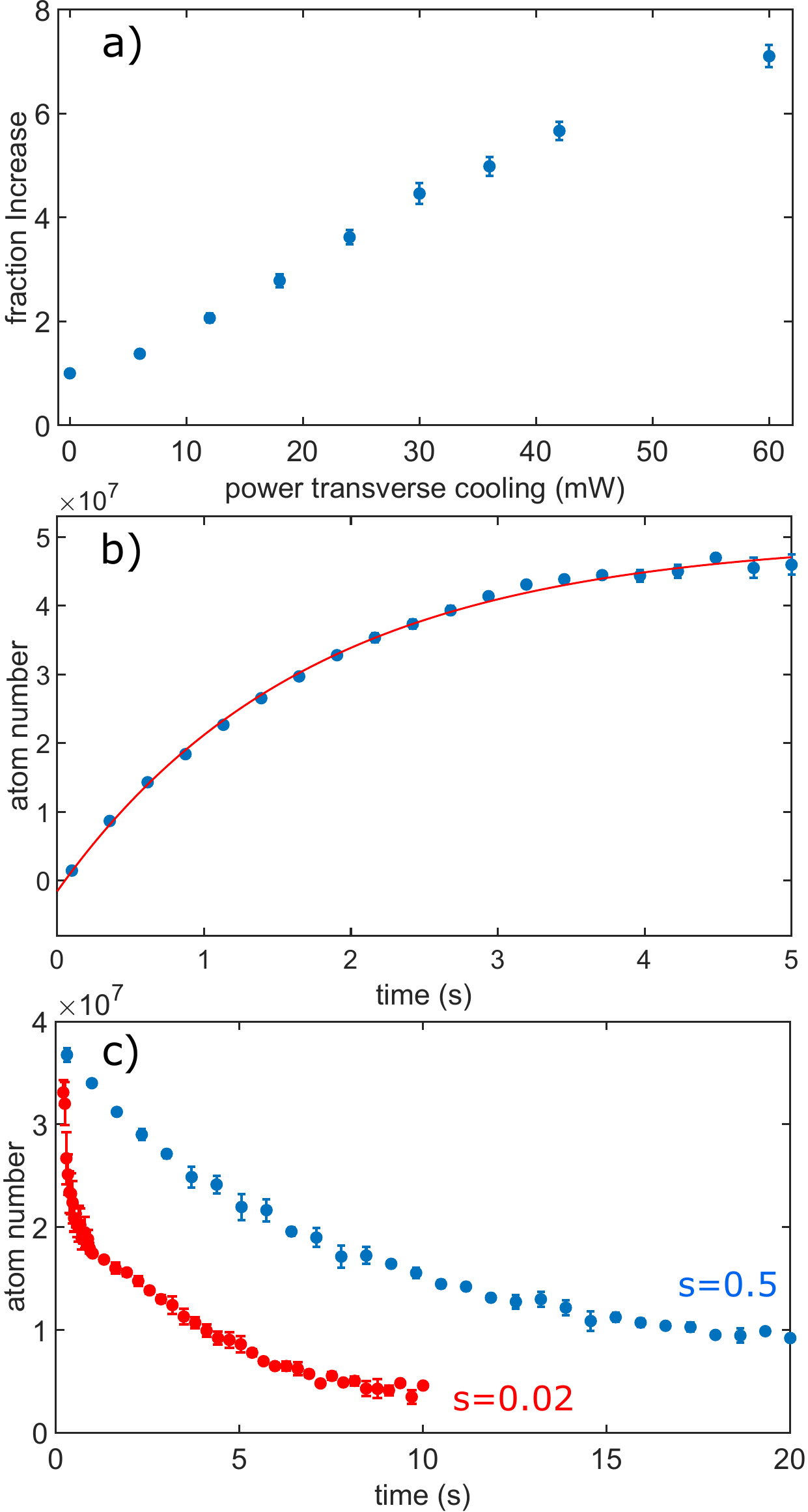}
\caption{\label{MOTs}MOT characterization. a) Increase of the atom number in the CMOT as a function of total laser power of the TC. b) MOT loading vs time. The atom number is measured after the CMOT stage at $s=0.02$. c) Atom number vs time in a CMOT for $s=0.5$ (blue) and $s=0.02$ (red).}
\end{figure}

Exiting the ZS the atoms are captured in a 626-nm narrow-line six-beam MOT~\cite{Frisch2012, Maier2014, Muhlbauer2018May}. We increase the MOT capture velocity by creating a comb structure of sidebands spaced by 110 kHz. We use a resonant EOM (Qubig PM6-VIS) to generate an effective linewidth of 7~MHz. During the MOT loading process (Fig.~\ref{MOTs}b), which takes 2~s, the laser detuning is $-38\Gamma_{626}$. We gain a factor of 2 in MOT loading rate ($2.85\times10^7$~s$^{-1}$) utilizing the above spectral broadening. Following the loading we enter the CMOT stage, where we reduce the temperature of the sample by extinguishing first the spectral broadening in 2~ms and successively ramp down the magnetic field gradient $1.5\rightarrow 1$~G/cm and the total laser intensity from $625\rightarrow 0.02I_s$ (saturation intensity $I_s=72~\mu$W/cm$^2$, MOT beams $\diameter$45~mm) in 50~ms. In parallel we linearly ramp down the laser detuning toward the resonance $\Delta_{626}=-38\rightarrow -10\Gamma_{626}$. This is not to be confused with the local detuning $\Delta_{\rm loc}$ relative to the $m_J=-8\rightarrow m_J=-9$ which remains constant at $-\Gamma_{626}/2(\sqrt{(\eta-2)s-1})$, where $\eta=\hbar k\Gamma_{626}/2m_\mathrm{Dy}g$, $m_\mathrm{Dy}$ is the mass of Dy, $g$ is the gravitational acceleration, $s$ is the saturation parameter and $k$ is the wavenumber for the 626-nm transition~\cite{Dreon2017}. At higher final laser intensities we achieve a higher atom number in the CMOT of $2\times10^8$ but at higher temperature. We further hold the sample for another 300~ms to reach a temperature of $5.6~\mu$K which is above the limiting Doppler cooling temperature of $T_{min}=\eta\hbar\Gamma_{626}/(2k_B(\eta-2))\simeq 3.4~\mu$K. The 300~ms hold time exceeds the time for the sample temperature to reach a plateau ($\sim$50~ms) but it is necessary for establishing local density equilibrium and therefore is needed for optimum dipole trap loading which follows as a next stage. 

To characterize the atom losses in the MOT we take two decay curves at a fixed detuning $\Delta_{626}=-10\Gamma_{626}$ and saturation parameter $s$ for two different final values of $s=0.02$ and $s=0.5$ (Fig.~\ref{MOTs}c). We make a leap forward and mention that we achieve the best phase-space density (PSD) in the dipole trap loading, following the CMOT stage, at $s=0.02I_s$. The PSD is calculated as PSD$=N(\hbar\bar{\omega}/k_BT)^3$, where $N$ is the atom number, $\bar{\omega}$ is the geometric average of the trap frequencies along orthogonal directions and $T$ is the temperature. Only for these extremely low $s$ parameter values we observe a steep initial decay, until a plateau with a very slow decay ($>20$~s) is reached (Fig.~\ref{MOTs}c, $s=0.02$). This initial decay is caused most probably by diffusion and loss of atoms from the MOT, which is unable to hold the complete cloud. However, in the next stage, when the MOT is supplemented with a dipole trap, this is when we get the best performance. The background pressure is $\simeq 1\times 10^{-11}$~mbar and therefore we disregard the slow one-body decay due to collisions with background atoms and attribute this decay to one-body scattering processes due to the 626-nm light. For larger saturation value of $s=0.5$ the observed initial loss can be attributed to two-body loss due to light-assisted collisions~\cite{Dreon2017} (Fig.~\ref{MOTs}c, blue trace). We extract a two-body loss rate from the initial slope of $\beta=(1/N\bar{n})dN/dt=4\times10^{-11}$~cm$^3$/s, where $\bar{n}=N/(8\pi^{3/2}\sigma_x^2\sigma_z)$ is the average density and $\sigma_{x,z}$ are the cloud size along $x$ and $z$ assuming $\sigma_x=\sigma_y$. The obtained loss rate is close to the theoretical prediction, based on a model of light-induced collisions, described in~\cite{Dreon2017}. We vary the two most important parameters to achieve the best loading conditions for the dipole trap, namely detuning $\Delta_{626}$ and $s$. For the temperature dependence on the saturation parameter we observe a linear slope of $T/\sqrt{s}=28~\mu$K for $s\in(0.5,10)$. We do not observe any temperature dependence on $\Delta_{626}$ for red detuned values $>2\pi\times 0.3$~MHz and fixed $s$, but a large dependence of the spin polarization. 
\begin{figure}
\includegraphics [height=5.2cm, width=8cm]{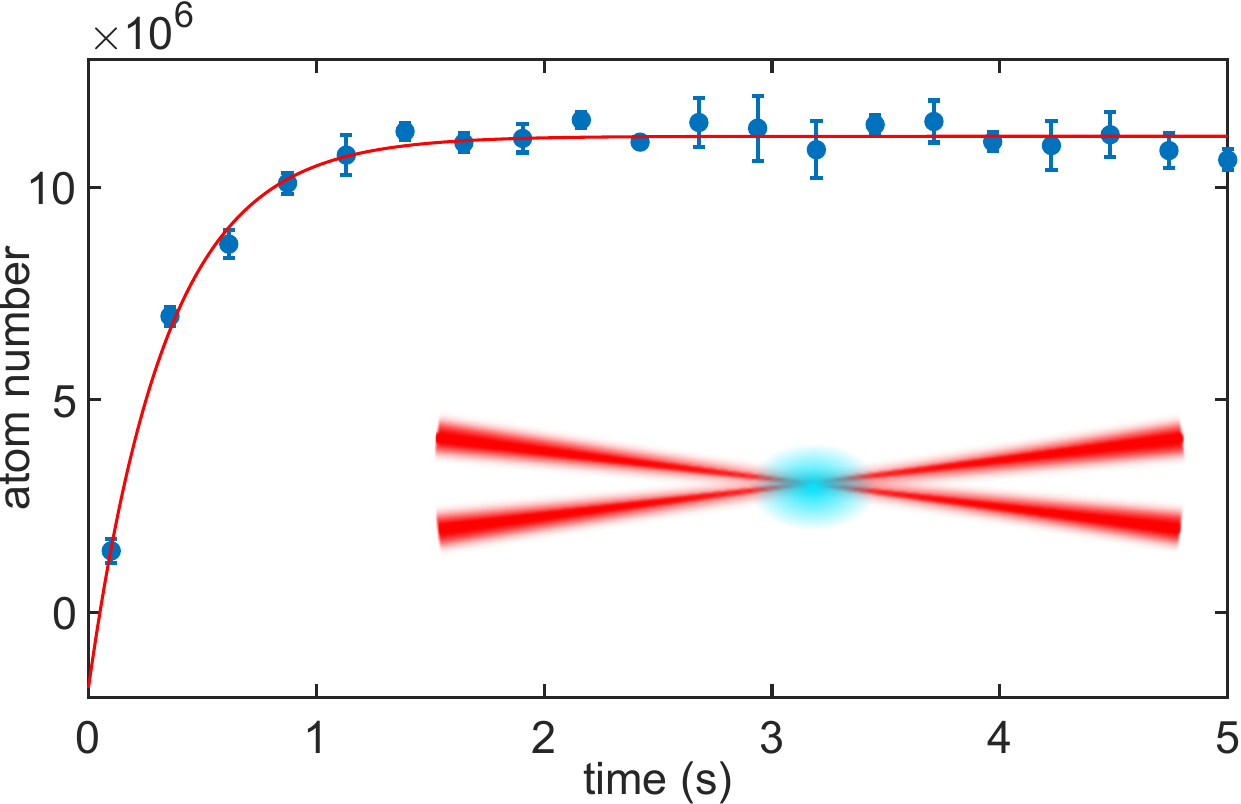}
\caption{\label{tweezers_loading}Tweezer traps loading vs time.}
\end{figure}

To quantify the degree of spin polarization we perform a Stern-Gerlach experiment after the MOT light is switched off, where we briefly turn on a gradient field before imaging. Given the size and the temperature of the MOT we cannot resolve the individual spin components. Instead we record the vertical cloud size after Stern-Gerlach expansion $\sigma_z^{\rm sg}$ and compare it to the bare size without gradient field $\sigma_z^0$, to get the normalized size $\sigma_z^n=\sigma_z^{\rm sg}/\sigma_z^{0}$, which we use as merit for the polarization purity. At detunings $\Delta_{626}\geq 2\pi\times1.4$~MHz, we find $\sigma_z^n\rightarrow 1$, meaning that the cloud is polarized. This is consistent with the behavior observed in other experiments~\cite{Dreon2017,Maier2014,Frisch2012}, explained by the sagging of the cloud under the quadrupole center in order to provide equilibrium between gravity and the radiative forces, subsequently leading to optical pumping due to the predominant $\sigma^{-}$ transition. We choose the limiting value of $\Delta_{626}=2\pi\times1.4$~MHz as a compromise between spin purity and cloud size, given that the horizontal size $\sigma_x$ grows linearly with the detuning and compromises the dipole trap loading efficiency later on. The spin polarization is crucial for the subsequent evaporative cooling in the dipole trap. 

Out of scientific curiosity, we performed eveporative cooling (described in Sec.~\ref{Evaporative cooling and transport}) starting with an unpolarized sample (using detuning $\Delta_{626}=2\pi\times 0.5$~MHz), which was possible only in a very narrow $<10$~mG magnetic field region around zero magnetic field. Apart from the technical difficulty, eventually we barely achieve the onset of a BEC with a tenfold compromised atom number. Eventually the sample becomes polarized during the evaporation ramp but in that process dipolar relaxation heats the sample and impairs the evaporative cooling efficiency.

\subsection{Tweezer loading, evaporative cooling and transport}
\label{Evaporative cooling and transport}
\begin{figure}
\includegraphics [height=7cm, width=8cm]{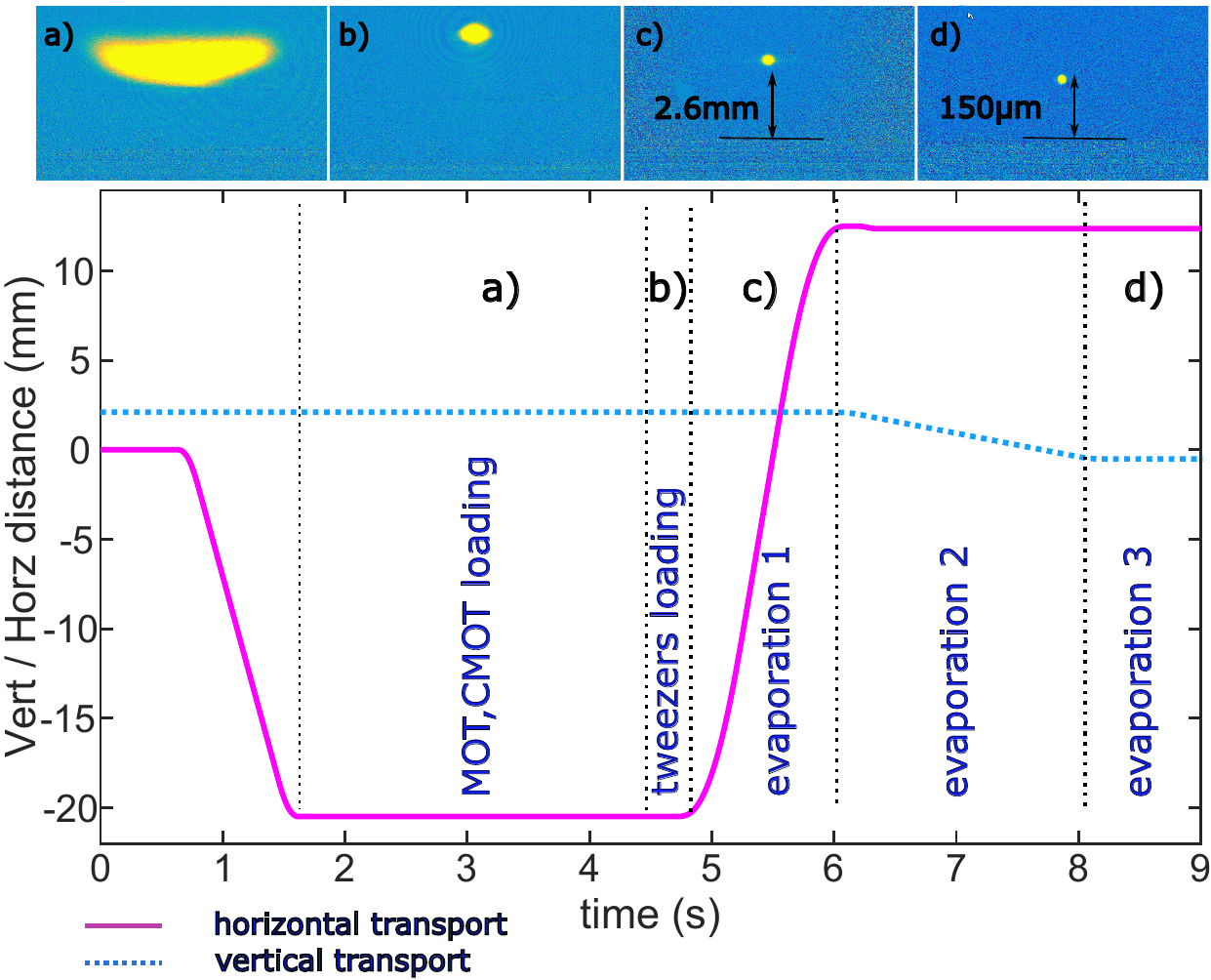}
\caption{\label{transport_pics}Horizontal and vertical transport with time. The accompanying MOT, CMOT and tweezers loading regions together with the three evaporation ramps are bounded by the vertical dashed lines (the state of the cloud on the pictures refers to the end of each period). Top row pictures are absorption images of Dy sample after a) 2.5~s MOT $+$ 300~ms CMOT, b) tweezers 100~ms after CMOT at aspect ratio 2.5 and $P$=4~W c) tweezers after horizontal transport and first evaporation down to $P$=2~W, d) tweezers after vertical transport and evaporation 2 and 3 down to $P=0.07$~W. Onset of a BEC appears after evaporation stage 2. Images a) and b) are taken with camera 1 along $x$ at the MOT region, and images c) and d) with camera 2 at the CAM region along $y^{\prime}$(see text).}
\end{figure}

In order to load the tweezers, they are already turned on during the MOT stage with a power of 11~W per beam. It is crucial that both beams have linear polarization in the $x-y$ plane (Fig.~\ref{transport_pics}). The difference between the $x-y$ and a $z$ polarization arrangement is a slight loading difference from the CMOT, possibly related to the tensorial part of the polarizability, and a much more dramatic difference in the atomic lifetime in the trap by an order of magnitude. Since the vertical position of the CMOT sensitively depends on the saturation parameter $s$, the detuning $\Delta_{626}$ and the vertical offset magnetic field $B_z$, but the first two are crucial for the final temperature and polarization of the sample, we choose to vary $B_z$ in order to match the CMOT vertical position with the tweezers tight vertical dimension. 
We increase the aspect ratio of the tweezer beams by a factor of 10 during the loading from the CMOT (at our optimum conditions of $s=0.02$ the CMOT sizes are $\sigma_z=0.2$ and $\sigma_{x,y}=1$~mm respectively). 
This dynamic mode-matching to the post laser-cooled cloud size improves our loading efficiency from the CMOT by a factor of 2.5 to a rate of $3.7\times 10^7$~s$^{-1}$ and an absolute number of trapped atoms of $1.3\times 10^7$ (Fig.~\ref{tweezers_loading}).

\begin{figure}
\includegraphics [height=5cm, width=8cm]{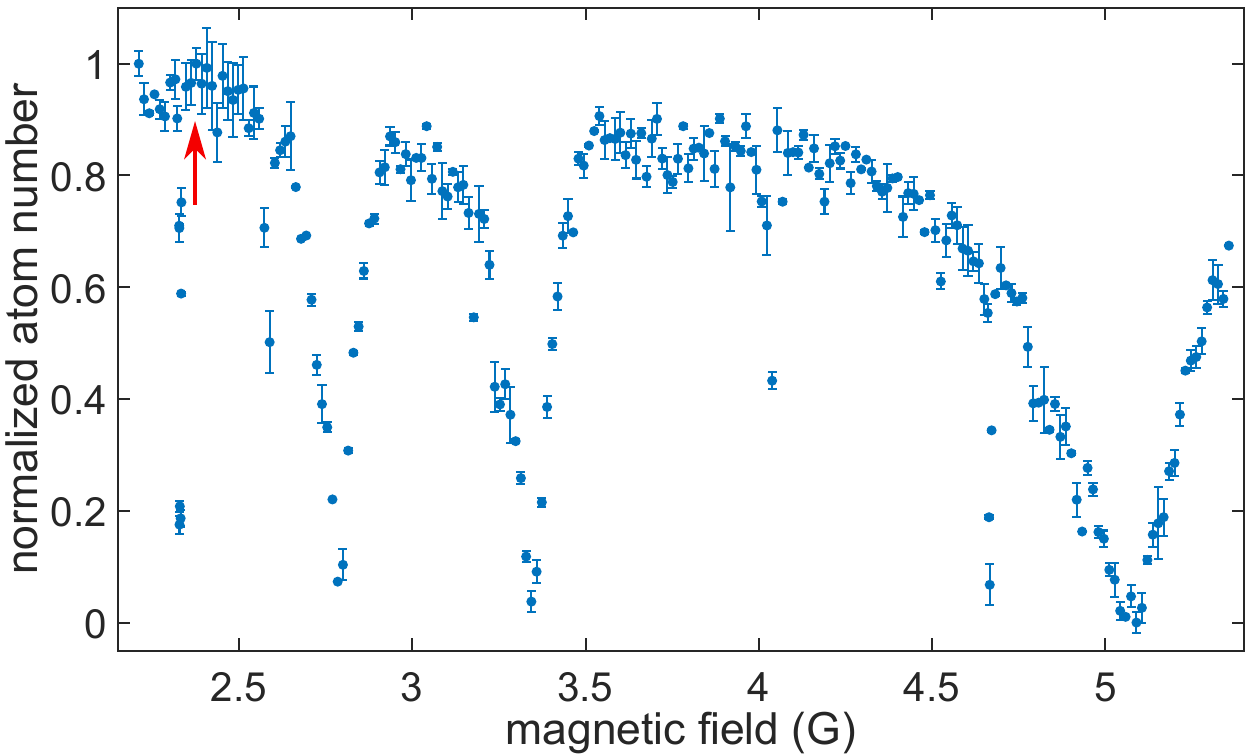}
\caption{\label{Fescbach_spectra} Atom-loss spectrum taken at an initial temperature of 200~nK in the low magnetic field range of interest. The arrow point at the region where the main evaporation and measurements in the manuscript are performed. The hold time at every point is 500~ms. Normalization is done relative to the atom number at zero holding time.}
\end{figure}

\begin{figure}
\includegraphics [height=5.1cm, width=8cm]{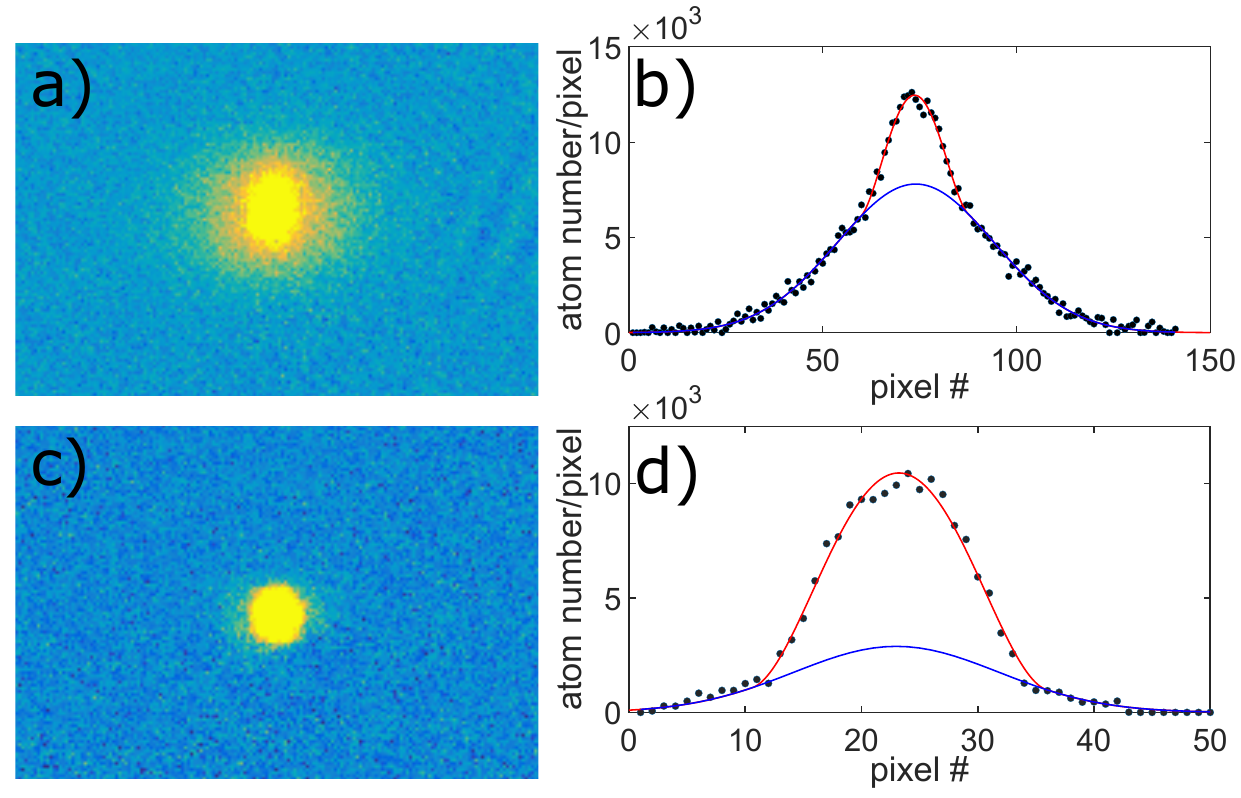}
\caption{\label{BECs} Phase transition to a BEC. Optical density profile a) and c) and 1D atomic densities b) and d) revealing bi-modal structure. Both bi-quadratic and thermal Gaussian fits are displayed in b) and d) to describe the BEC fraction and thermal component respectively. Atom number, temperature and BEC fraction are a), b) 4.5$\times 10^5$ atoms, 200~nK, 15$\%$ and c), d)  1.5$\times 10^5$ atoms, 40~nK, 70$\%$. The effective pixel size in the object plane is 4~$\mu$m. The time-of-flight is 25~ms. The imaging is performed in absorption with camera 2 along the $y^{\prime}$-axis.}
\end{figure}

After the tweezer loading, we extinguish the modulation and linearly ramp the powers of the two tweezers to 2~W each in 500~ms.  At this point we have a $\mathrm{PSD}=0.007$ and $2.5\times 10^6$ atoms at $9~\mu$K (trap frequencies $(f_x, f_y, f_z)=(51, 398, 874)$~Hz). Next we initiate the first out of three evaporation ramps, and simultaneously the horizontal transport. As shown in Fig.~\ref{transport_pics} (magenta solid trace) we transverse the horizontal gap between the MOT and CAM regions in 1~s and exponentially lower the power of the tweezers to $P=0.5$~W (atom number $9\times 10^5$, $T=2~\mu$K, $\mathrm{PSD}=0.03$, trap frequencies $(f_x, f_y, f_z)=(26, 198, 437)$~Hz). Until this point we evaporate at the same magnetic field of $B_{0}=0.8$~G as the one set during the CMOT stage after extinguishing the CMOT gradient field. We set this field with our large offset coils ensuring a good spatial uniformity during the transport. Once the cloud is transported above the CAM we engage also the local offset coils, concentric with the CAM (Fig.~\ref{transport_cross_section}), which allows us to adjust the $z$-field in the foothills of the Feshbach resonance centered at 2.75~G (Fig.~\ref{Fescbach_spectra}). The field is switched to the new value of 2.5~G (scattering length $\sim$120a$_0$) in 5~ms and we initiate the vertical transport (Fig.~\ref{transport_pics}, blue dashed trace). The same Feshbach region is mapped with the magnetic field pointing along $y$-direction. Rotating the magnetic field in a precise manner, while maintaining a particular contact scattering length is important for future plans to achieve a 2D sample by effectively canceling the DDI and the contact interaction (see~\ref{Conclusions and Perspectives}). During the vertical transport, which takes 2.5~s, we execute the second evaporation ramp ending at $P=0.12$~W. We see the onset of a BEC at the end of evaporation 2 at $T=200$~nK (Fig.~\ref{BECs}a) and b)). The last ramp is executed in 1~s and delivers a BEC with a condensate fraction of 70$\%$ (Fig.~\ref{BECs}c) and d)). During the time of flight the DDI preserves the parabolic condensate profile~\cite{Santos250403} and although the Thomas-Fermi criteria is changed when DDI is added to the conventional $s$-wave interaction, we satisfy the extended criteria for our parameters ~\cite{Parker2008}.

During the last ramp we have the option to reshape the cloud by including in the evaporation either a vertical broadband 770-nm beam ($\diameter 150~\mu$m), or two orthogonal horizontal 1053-nm dipole traps ($x\times z~\diameter=300\times 70~\mu$m$^2$ each) that coincide with the x$^{\prime}$y$^{\prime}$ coordinate system (Fig.~\ref{transport_cross_section}). The horizontal traps can be turned into lattice beams later in the experimental sequence (detailed below). The automatic tuning of the aspect ratio of the tweezers completes the full control over all the trap frequencies. The broadband 770-nm dipole trap goes in the negative $z$-direction through the CAM. The undesired lattice formation from imperfect AR coatings of the CAM optical surfaces is evaded by the short coherence length  $l_c=2ln2\lambda_{770}^2/\pi\Delta\lambda_{770}=29~\mu$m of this laser source relative to the distance between the cloud and these surfaces.

\subsection{Optical lattices characterization}
\label{Optical lattices}
\begin{figure}
\includegraphics [height=5cm, width=8cm]{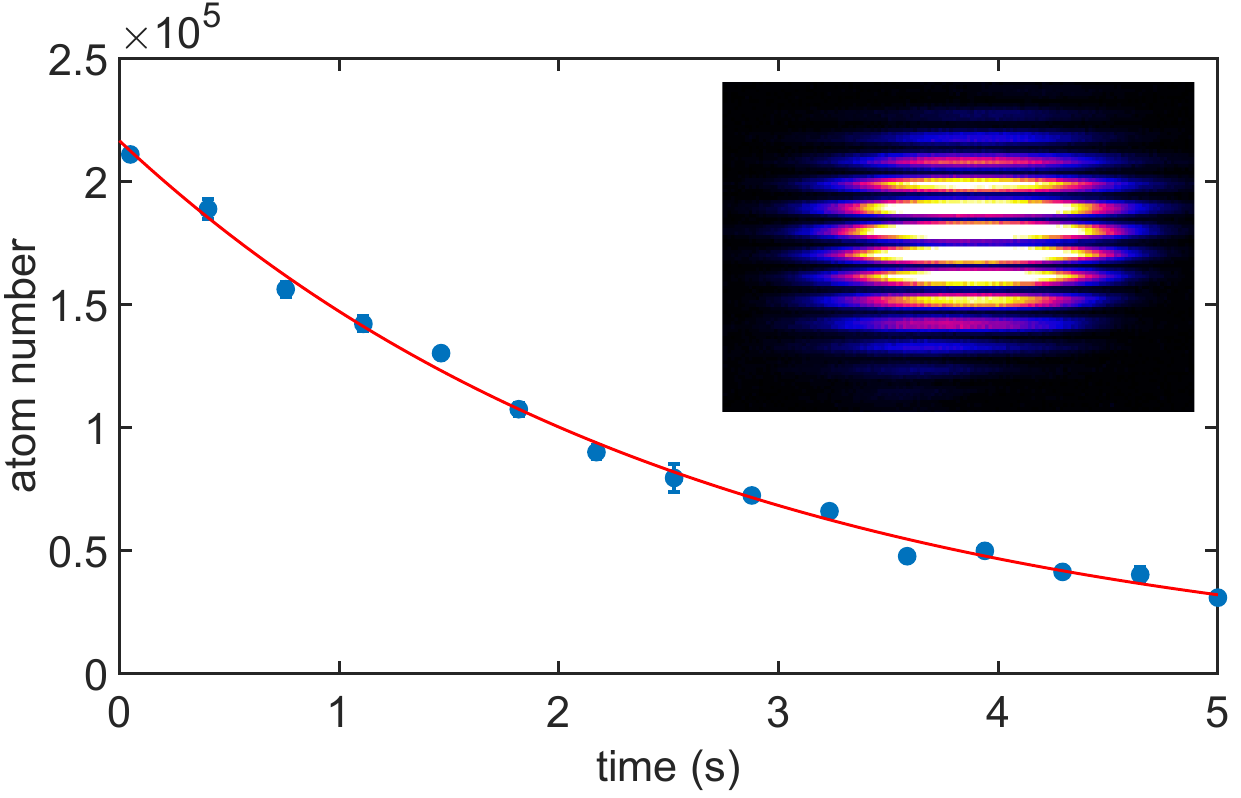}
\caption{\label{accordion_decay_pic} Lifetime of the Dy cloud loaded in the 532-nm FA with a trap frequency $\omega_z=2\pi\times1$~kHz. The inset shows the FA interference pattern.}
\end{figure}

The inset of Fig.~\ref{accordion_decay_pic} displays the interference pattern of the FA with $\sim$13$~\mu$m spacing, imaged by a CDD. We load the atomic cloud in a single fringe by simultaneously raising the FA power and the 772-nm vertical beam power in 500~ms, so we can barely hold the cloud against gravity. Meanwhile we linearly ramp down the tweezers power to zero. Figure~\ref{accordion_decay_pic} shows an exponential decay of the atom number with time constant of 2.6~s in a FA with $\omega_z=2\pi\times1$~kHz trap frequency along the $z$-direction. The atom number is deduced by capturing fluorescence on the 421-nm atomic transition for 1~$\mu$s with the CAM. The polarization of each of the FA beams is in the $x\mbox{-} z$ plane. The lifetime is reasonable for future experiments, and we don't observe a detectable shift from the 421-nm resonance, while holding the atoms in the FA during imaging. We can reach up to $\omega_z=2\pi\times4$~kHz vertical trapping frequency and at the typical radial confinement determined mostly by the 770-nm trap of $\omega_{\rho}=2\pi\times54$~Hz we satisfy the criteria for the 2D mean-field regime $Na_s(1+2\epsilon_{dd})l^3_z/l^4_{\rho}\sim 0.2\leq 1$ (for atom number $N=10^4$, $s$-wave scattering length $a_s=88a_0$, ratio of the dipolar to $s$-wave scattering length for $^{164}$Dy $\epsilon_{dd}=a_{dd}/a_s=1.5$, and oscillator lengths in the tight confinement and radial direction $l_z$ and $l_{\rho}$, respectively)~\cite{Parker2008}. The expansion after releasing from the FA is anisotropic and in the vertical waist follows the expected linear time dependence of $R(t)=\sqrt{3.5\hbar\omega_0/m_\mathrm{Dy}}t$, where $\omega_0$ is the trap frequency along the tight vertical $z$-direction and $t$ is the time of flight~\cite{Ketterle2001}. 

\begin{figure}
\includegraphics [height=12.4cm, width=8cm]{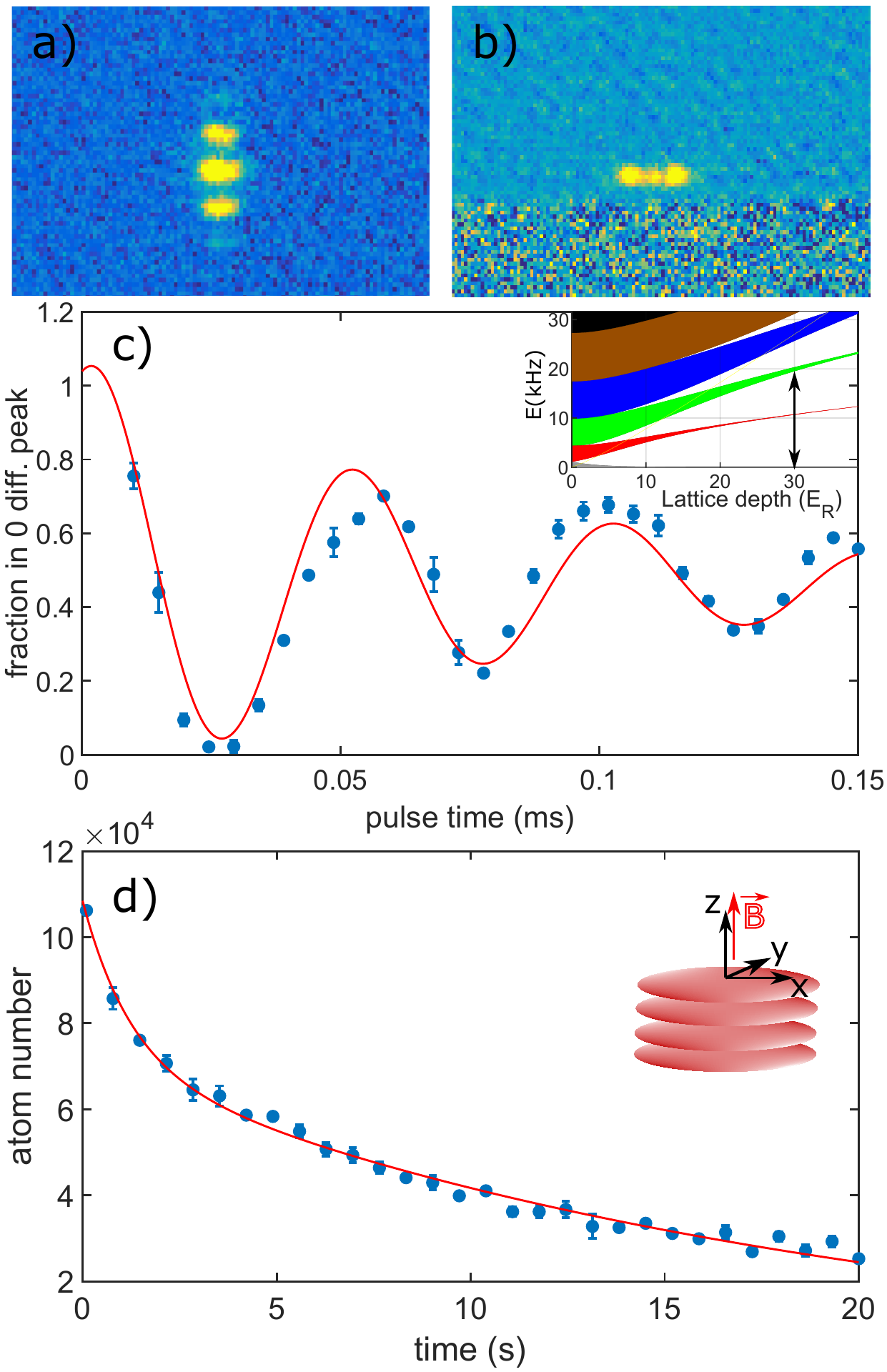}
\caption{\label{all_lattice} Characterization of optical lattices. Population in the $0\hbar k$ and $\pm 2\hbar k$ momentum states for both vertical 1053-nm lattice a) with 10~ms time-of-flight, and a horizontal $x^{\prime}$ lattice beam at 532~nm b) with 4~ms time-of-flight. The boundary visible on panel b) represents the CAM front surface.  c) A lattice depth measurement showing oscillations between $0\hbar k$ and $\pm 2\hbar k$ momentum states after sudden loading in a lattice. The inset shows the 0-2 band gap at at E$_R$. d) Lifetime of the full cloud (BEC and thermal fraction) loaded in the vertical retro-reflected from the CAM lattice at 1053~nm.}
\end{figure}

To calibrate the short lattices at 1053~nm and 532~nm, as described in Sec.~\ref{Optical lattices implementation}, we use a standard Dirac-Kapitza diffraction method, where we pulse the lattice power ($<0.5~\mu$s rise/fall time) over a short period and vary its time duration~\cite{Ovchinnikov1999,Denschlag2002}. The actual momentum peaks of the matter wave interference pattern are shown as examples in Fig.~\ref{all_lattice}a,b, for both horizontal 532-nm and vertical 1053-nm lattices. We detect oscillations between the population in the $0$ and $\pm2\hbar k_{1053}$ momentum states ($k$ is the wave-vector for the 1053-nm lattice), with a frequency corresponding to the energy difference between the zeroth and second energy bands. An example of such oscillation in the vertical 1053-nm lattice at $\sim 30E_R$ ($E_R=h^2/(2m_\mathrm{Dy}\lambda_{1053}^2)$ is the photon recoil energy, $h$ is the Plank constant and $\lambda_{1053}$ is the wavelength at 1053~nm) is shown in Fig.~\ref{all_lattice}c). We have calibrated similarly our 532-nm lattices. The decay of the oscillation is possibly due to the inhomogeneity of our lattice beams and the presence of both contact interaction with scattering length of about $100a_0$ and DDI. The lifetime of a sample adiabatically  loaded into the vertical lattice, at powers barely holding the atoms against gravity, reveals a dual exponential decay with the longer lifetime $>20$~s (Fig.~\ref{all_lattice}d). 

\section{Fluorescence imaging with the CAM}
\label{Fluorescence imaging with the CAM}
Typically CAMs with alkali atoms~\cite{Gross2021Dec}, termed quantum gas microscopes (QGM) when they operate with a single site resolution, are supplemented with a cooling mechanism that keeps the atoms pinned in the lattice as they fluoresce. Currently we are attempting another approach given the large mass of Dy and the broad 421-nm imaging transition. 
The scheme is similar to the one implemented for $^6$Li in~\cite{Bergschneider2018}, based on a collection of few photons per pixel in a short time. Given our significant overall collection efficiency of approximately 20$\%$, we are currently pursuing a fast illumination of $\tau =1-2~\mu$s during which we detect 20-40 photons/atom, or $\sim 2-4$ photons/pixel/atom given our current magnification of 160~\cite{Miranda2015}. During fluorescence, the atoms will exhibit diffusion from photon recoil. As a result, the width of the position distribution scales as $\propto\tau^{3/2}\Gamma_{421}^{1/2}/m_\mathrm{Dy}\lambda_{421}$~\cite{Joffe1993,Bergschneider2018}. Given the large mass of Dy, for times $<3~\mu$s the atom will not exceed the width of a single lattice site of a 1053-nm standing wave, and therefore one can even turn off the lattices at the time of the imaging.
For this preliminary work lattices and traps are always kept on during the fluorescence detection, which additionally inhibits their random walk.

In detail our overall collection efficiency is defined mostly by the collection efficiency of the CAM of 30$\%$, which accounts for the photon angular distribution, based on linear polarization of the excitation beams in the $x-y$ plane and the applied magnetic field along the axis of the microscope. 
The total transmission of the CAM is further limited by the quality of the AR coating ($R<0.3\%$ per surface) and two narrow transmission filters centered at 420-nm (Semrock, FF01-420/10, $T>95\%$) placed in front of camera 3. Furthermore the fluorescence is guided by two dichroic mirrors (Semrock, Di03-442-t3, $R>97\%$) in reflection to the camera, which has quantum efficiency of $83\%$.

\begin{figure}
\includegraphics [height=2cm, width=8cm]{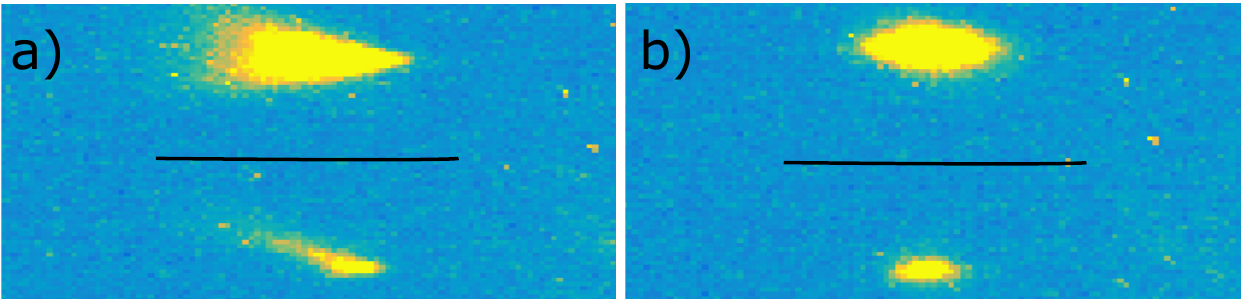}
\caption{\label{beam_compensation} Compensation of excitation beam radiative force effects. a) Side fluorescence imaging of Dy cloud showing an accelerated Dy cloud due to unbalanced excitation beam. b) The same as a) with balanced excitation beam by counter-propagating it with orthogonal polarization. The black line in a) and b) shows the surface of the CAM. Underneath, the cloud mirror image from the top CAM surface is visible.}
\end{figure}

For the above strategy to succeed we first need to eliminate the linear net radiative force on the atoms by a pair of counter-propagating excitation beams. Each beam is resonant with the 421-nm Dy transition, with an intensity $>100I_{421}$, and its focus at the atom position is $\diameter 150~\mu$m. The tight focus avoids clipping from the front face of the CAM. The pair is practically co-propagating with the $y^{\prime}$ lattice, making only a $3^{\circ}$ angle with it, which allows us to deviate the excitation beams from the lattice light after the cloud. However, this will tilt the lattice potential along this direction. As an example, the needed lattice depth for a 532-nm lattice to contain the atoms against this tilt, if a single beam was used, is $V_{min}=\hbar k_{421}\Gamma_{421}/2k_{532}\sim 1.9$~mK. Here $k_{421}$ and $k_{532}$ are the wave-vectors of the excitation and lattice light at 421~nm and 532~nm, respectively. This is rather substantial given the power available for our lattices. Currently we strobe each beam with 0.5-$\mu$s long pulses, ensuring the  counter-propagating pulse trains are out of phase~\cite{Bergschneider2018, Su2024}. The strobing is implemented by a two-channel arbitrary waveform generator (Keysight 33509B) activating two independent AOMs, gating the light of each addressing beam. The tight beam focusing at the AOM crystal allows pulses as short as $0.1~\mu$s. 

\begin{figure}
\includegraphics [width=\columnwidth]{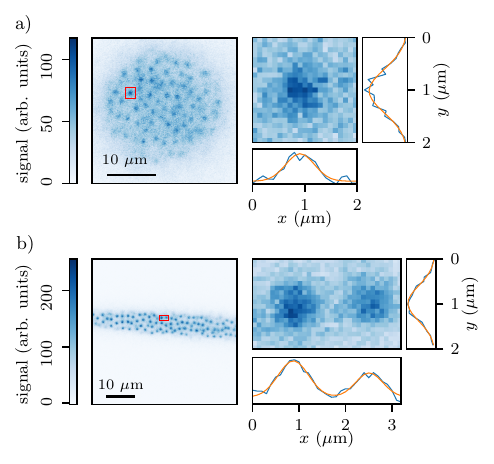}
\caption{\label{CAM_pics} Fluorescence imaging with the CAM of a multi-droplet crystalline excited state in: a) pancake-like configuration at trap frequencies $(f_x, f_y, f_z)=(11, 76, 160)$~Hz; and  b) cigar-shaped configuration with trap frequencies (26, 24, 124)~Hz. The zoom-in on the right reveal minuscule droplets of a few hundred atoms with a diameter of $0.6~\mu$m. Illumination time is 2$\times 1~\mu$s}
\end{figure}

To assure a good overlap with the original illuminating beams, we utilize fluorescence imaging transversely to the CAM using camera 2. We first block one of the beams and provide a single excitation pulse of $50~\mu$s. The fluorescence image resembles a cone aligned in the direction of the excitation beam, formed by atoms being accelerated along its $k_{421}$ wave-vector (Fig.~\ref{beam_compensation}a). We then carefully align the second beam until we see no offset in the original position of the cloud (Fig.~\ref{beam_compensation}b). The leftover heating during the excitation is due to diffusion. The expected heating rate due to momentum kicks of absorption and emission is $dT/dt=\Gamma_{421}\hbar^2k_{421}^2/2k_Bm_{\rm Dy}=67~\mu$K/$\mu$s. 

\begin{figure*}
\includegraphics [width=\textwidth]{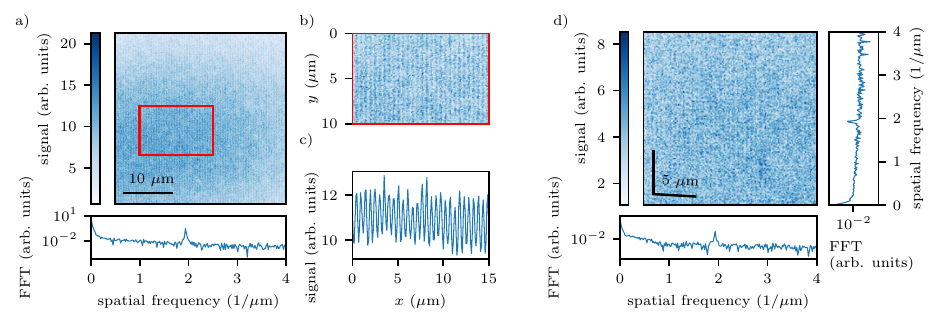}
\caption{\label{insitulattices}$In\:situ$ images of a 1D a) and 2D d) optical lattice with 526-nm spacing, taken with imaging time 2$\times$1$~\mu$s and magnification of 160. Camera pixel size is 16~$\mu$m. The spatial Fourier transform peaks at spatial frequency $\sim 1.9~\mu$m$^{-1}$, which fits the inverse lattice spacing. b) shows the red cutout in a) and c) the corresponding signal averaged along the vertical direction, revealing the periodic density modulation in the lattice. d) The angle between the two lattice directions deviates by $3.3^{\circ}$. The Fourier spectra are computed by rotating the image correspondingly.}
\end{figure*}

Once the addressing 421-nm beam is balanced, we adjust the sample with the tweezers at a specific height above the CAM, in the vicinity of its focal plane. To navigate ourselves, if we are within 1-2$~\mu$m of the focal plane we need to create a mask of point sources or stripes, created by the atoms, and use their contrast as a metric for this initial vertical positioning. We take advantage of a peculiar droplet crystalline state~\cite{Blakie2016}, consisting of multiple sub-$\mu$m droplets (Fig.~\ref{CAM_pics} a,b). This state is achieved by a sharp quench (on a few-ms scale) from a BEC state which we evaporate into at a typical magnetic field of $B_{in}=2.5$~G and scattering length of $\approx$120$a_0$~\cite{Tang2015,Chomaz2019Apr}, into the region of an insulating droplet state at final magnetic field of $B_f=0.7$~G ($\approx$90a$_0$). This droplet crystalline state differs from the ground state at the final point of the quench, which based on a simulation using an extended Gross-Pitaevskii equation (eGPE)~\cite{santos2016Jun,Barbut2016,Bisset2016,Chomaz2016Nov,Norcia2021,Bland2022} and on an $in\: situ$ fluorescence image with the CAM contains only a few large droplets. The excited crystalline state (lifetime $>100$~ms) is therefore reminiscent to a high resolution microscopy slide target. We use this method for different final cloud aspect ratios. To obtain a cigar-shaped cloud we evaporate solely in the two-tweezer configuration with anisotropic frequencies $(f_x, f_y, f_z)= (11, 76, 160)$~Hz and atom number of 1.5-2$\times 10^5$ (Fig.~\ref{CAM_pics}b). Alternatively we prepare a pancake-like configuration with trap frequencies $(26, 24, 124)$~Hz (Fig.~\ref{CAM_pics}a), where we use our $x-y$ lattice beams as traps (see Subsec.~\ref{Optical lattices implementation}) by introducing them in the final evaporation step and simultaneously dynamically increase the horizontal waists of the tweezers by a factor of 3 with our modulation method.     

Once in the vicinity of the CAM focal plane we prepare a 2D sample by ramping up the accordion vertical lattice and simultaneously the axial time-incoherent 770-nm trap in 100~ms (typical trap frequencies $(f_x, f_y, f_z)=(75, 75, 1.2\times 10^3)$~Hz). Beforehand, the sample is squeezed and mode matched to this 2D geometry by linearly ramping up the tweezers and simultaneously dynamically increasing their horizontal waists (by a factor of 5.3), achieving trap frequencies of $(f_x, f_y, f_z)=(62, 46, 540)$~Hz. This is followed by ramping down the tweezers to zero power in 100~ms and leaving the sample purely to the accordion-axial 770-nm trap. For these experiments we used the interferometer-based accordion and therefore the vertical position of the atoms was adjusted by tuning the piezo actuator that changes the length of one of the interferometer arms. 

Now we are in position to raise the horizontal $x^{\prime}-y^{\prime}$ lattices. We choose an exponential ramp in 100~ms, while we additionally raise the accordion power to achieve a vertical trapping frequency of 3~kHz and a corresponding oscillator length of 140~nm. In Fig.~\ref{insitulattices} $in\: situ$ images of a single 1D and a 2D lattice with spacing of 526~nm are displayed, taken with $2\times 1~\mu$s total imaging time. For these images we limit the atom number to $<10^4$. To minimize the axial defocus, we compute the spatial Fourier transform of the grid of atoms loaded in a 1D lattice, and use the peak at a spatial frequency of $1.9~\mu$m$^{-1}$ (inverse lattice spacing) as a figure of merit. The two lattices deviate from orthogonality by $3.3^{\circ}$. Currently we observe slightly worse contrast in the lattice $y^{\prime}$ probably due to imperfect balancing of the two counter-propagating imaging beams, which are almost aligned with the $y^{\prime}$ direction. 

Being able to resolve single lattice sites with high fidelity, eventually even at lattice spacings similar to or below the resolution of our CAM, driven by the necessity to achieve higher DDI rates, will requite the introduction of new tools. Generally one needs to accommodate a few competing factors: On one hand to have high resolution in the image plane  requires a large magnification due to the large pixel size of the iXon-897 of $16~\mu$m. Subsequently the photon count per pixel is reduced and one resorts to longer imaging times. In order to avoid active cooling during the imaging process, which makes this method appealing, the imaging time has to be limited such that the detection fidelity is not compromised due to the stochastic diffusion that will eventually overlap the photon signals of two neighboring sites. Magnifying the atom positions before the imaging is therefore the best approach, as demonstrated recently by the introduction of a magnifying accordion lattice for Erbium~\cite{Su2023, Su2024}. We describe our current strategy to resolve the above limitations which differ from other approaches in Sec.~\ref{Conclusions and Perspectives}. The optimal spacing that can still give a large available field of view and substantial detection fidelity is similar to the average distance between the droplets in our droplet crystal (Fig.~\ref{CAM_pics}).

\section{Conclusions and Perspectives} 
\label{Conclusions and Perspectives}
We presented a potential experimental platform capable of simulating spin models of quantum magnetism based on highly-magnetic Dy atoms pinned on an optical lattice. Additionally, Dy atoms prepared in a specific set of degenerate magnetic sublevels, members of a highly electrically coupled opposite parity doublet, cast a peculiar pseudospin-1/2 system, that features both eDDI and mDDI. Due to the degeneracy, the spin-orbit coupling term intrinsic to the DDI is activated additionally to the conventional Ising and spin-exchange (state swapping) DDI terms primarily considered. Unlike those terms the spin-orbit coupling changes the spin angular momentum by $\pm2$ and therefore imprints a phase factor of $\exp(\pm 2i\phi_{ij})$ associated with a change of the orbital angular momentum among an atom pair. 

Once the DDI is projected onto the above pseudospin-1/2 basis and rewritten in terms of a conventional spin-1/2 operator set $\hat{S}_{x,y,z}$ one arrives at a generalized $XYZ$ spin model where the asymmetry of the rates governing the $\hat{S}_x\hat{S}_x$ and $\hat{S}_y\hat{S}_y$ interactions is assured by the spin-orbit DDI. Furthermore these rates depend on the lattice geometry through the azimutal phase $\phi_{ij}$. Quantum simulators, such as this one, present an irreplaceable tool for exploring tunable low-symmetry models that could not be approached analytically or numerically.

Furthermore, we described a new apparatus and its basic operation, tailored to achieve the above goals. The apparatus consists of many atypical, or out of the mainstream alternatives to conventional cold atom experimental approaches, which we hope this work validates. Specifically, utilizing a large glass cell with an indium seal eliminates the necessity of a long glass to metal transition. Having a single cell serving as a main laser cooling vessel and containing a CAM facilitates the atomic transport. The glass cell nano-coating provides lossless access for any laser wavelength from the mid-UVC region to the NIR even at large angle of incidence. The CAM simplistic miniature design, allows its positioning in-vacuum, and therefore possesses a large NA, field of view and reasonable working distance, while eliminating the need to include glass vacuum ports into the optical design. The latter approach suffers from the glass warping due to the atmospheric pressure and the viewport manufacturing process itself. 

We have additionally complemented the vertical 1053-nm lattice, currently retro-reflected from the CAM front face with a secondary retro-reflected beam at 1095-nm. The formed beat-note lattice~\cite{Masi2021} should form the same long wavelength potential envelope as we currently have with the 532-nm FA (with step 13$\mu$m), with the difference that this potential will be referenced to the CAM. Our strategy in selecting a single reproducible 2D plane of atoms referenced to the CAM and coincident with its focal plane is to adjust the beat-note envelope such that the $n$-th anti-node coincides with the CAM focal plane. For that purpose the seeding laser of the 1095-nm fiber amplifier (Azurlight, 8~W) is an AR coated ECDL with broad tuning capability, which is needed due to the proximity of the retro-reflective mirror to the atoms (in our case the front surface of the CAM) of 160~$\mu$m. On the other hand the robustness of the beat-note envelope is assured even if the lasers run freely. Once we hand out the 2D sample from the FA to the beat-note lattice anti-node, even if the sample occupies few neighboring planes, we can attempt selective parametric heating by exciting the atoms in all planes to the second band gap except the plane of the anti-node. For this purpose the frequency of the parametric heating should be scanned and the scan interrupted right before the trapping frequency gap of the anti-node plane (which is about 1.5$\%$ higher than the next nearest neighbor). The beat-note lattice could be further used to load two nearby 2D layers allowing the engineering of bilayer dipolar systems~\cite{Pikovski2010,Safavi-Naini2013, Guijarro2022}. 

Furthermore we have nearly implemented a 2D dynamic accordion lattice (DAL)~\cite{Wili2023}, that can change in 30~ms the site spacing from 266~nm to 1.5$~\mu$m. The starting spacing is commensurate with the 532-nm science lattice spacing and with proper adjustment of its phase one could transfer the atoms in it without significant heating. This should allow a magnification of the 2D sample, before imaging at high fidelity, as recently demonstrated for Er in~\cite{Su2023}. 

Another elegant method that could be adapted to our situation, is currently also pursued in order to increase the imaging fidelity. The method is based on coherent expansion in a combination of a crossed horizontal 2D light sheet (reminiscent to the current FA) and a radial harmonic potential (our time-incoherent 770-nm dipole trap). Evolving the system up to a specified time slightly exceeding a quarter period of the harmonic potential, followed by precisely timed free expansion, could magnify the state in position space and access the $in\: situ$ distribution of the atoms, as it has been demonstrated for multiple occupied lattice sites~\cite{Asteria2021}. In the last work a large magnification $>90$ was achieved. Since we need a magnification of $<6$, to render the single atoms spacing of $\sim 1.5-2~\mu$m, we need to still be in the same proximity to the CAM focal plane. Therefore the free expansion in our implementation of the technique should be substituted with a expansion in a 2D plane with very low orthogonal trap frequencies.

\begin{acknowledgments}

We are highly indebted to R.N.\ Bisset for providing us with a eGPE Matlab simulation of the droplet formation and beat-note lattice loading. We thank M.\ Fattori for discussion regarding the beat-note lattice and its implementation and M.\ Lepers for providing us with details on the OPS. We thank C.\ Ravensbergen for providing the ZS design. We thank M.\ Sohmen and A.\ Jesacher for discussing possible improvements of the fluorescence detection utilizing active light tailoring devices. We thank A.\ Canali and C.\ Baroni for useful suggestions on this script. We acknowledge a generous financial support by an ESQ discovery grant (306532) and by the Julian Schwinger foundation (JSF-22-05-0010). 
\end{acknowledgments}

\appendix
\section{Detailed preparation of a degenerate isospin-1/2 system with mDDI and eDDI}
\label{isospin detail}
The level structure of the OPS can be seen in Fig.~\ref{levels_updated}. We depict the rightmost magnetic sublevels of the OPS manifold, as $|1\rangle=|J=10,m=-10\rangle$ (we drop the $J$ and $m$ notation from here on), $|\alpha\rangle=|10,-9\rangle$ and $|\beta\rangle=|9,-9\rangle$. We restrict ourselves to Dy atoms in the $x-y$ plane, with applied magnetic field   $\vec{B}=B\hat{z}$ and a MW $\vec{E_m}=E_me^{i\omega t}\hat{z}$ with linear polarization directed along the $\hat{z}$ direction. The MW $^{\prime\prime}$dresses$^{\prime\prime}$ the bare states $|\alpha\rangle$ and $|\beta\rangle$~\cite{Will2016, Karman2022}.

For this geometry and choice of states, the state $|1\rangle$ has no $^{\prime\prime}$partner$^{\prime\prime}$ in the $J=9$ to be coupled by the MW. Moreover the $g$-factors of both $J=9$ and $J=10$ states are almost identical ($g_{10}=1.3$, $g_9=1.32$) therefore we can consider the MW detuning $\delta$ as unaffected by the Zeeman shift. Then for certain magnetic field $B$, and MW detuning $\delta$ and Rabi frequency $\Omega$ we can induce degeneracy between the bare state $|1\rangle$ and the MW dressed state $|2\rangle=a|\alpha\rangle+b|\beta\rangle$. Here we defined the weights $a=\sin{\theta}$ and $b = \cos{\theta}\exp(i\omega t)$, with $\cos{2\theta}=-\delta/\sqrt{\delta^2+\Omega^2}$. The crossing will occur at magnetic field $g\mu_BB=\hbar/2(-\delta+\sqrt{\delta^2+\Omega^2})$ with the induced electric dipole moment in the $|2\rangle$ state of $\langle 2| d|2\rangle=(1/\sqrt{210})\langle10||d||9\rangle(1/\sqrt{1+(\delta/\Omega)^2})$. 

We now focus on the degenerate two level system ($|1\rangle$, $|2\rangle$) in the limit of small $\theta$, coupled by both mDDI and eDDI. In the geometry discussed, the relevant parts of the electric and magnetic DDI are $\hat{H}^2_0=(1/R^3_{ij})((\hat{d}^i_1\hat{d}^j_{-1}+\hat{d}^i_{-1}\hat{d}^j_{1})/2+\hat{d}^i_0\hat{d}^j_0)$ and $\hat{H}^2_{+2,-2}=(-3/2R^3_{ij})(\hat{d}^i_1\hat{d}^j_{1}e^{-2i\phi_{ij}} + h.c.)$. Here $\phi_{ij}$ is the azimuthal angle of the vector connecting the atoms, positioned at lattice sites $i$,$j$, $R_{ij}$ is the distance between them, and the polar angle is automatically substituted with $\pi/2$; superscripts refer to the rank of the spherical tensor and the subscripts to its components. Like in proposals based on magnetic atoms~\cite{Patscheider2020} or polar molecules~\cite{Gorshkov2011,Gorshkov2013}, when both mDDI and eDDI are projected in the $|1\rangle$,$|2\rangle$ basis one generates a spin model of the type $\hat{H}_\mathrm{spin}=\Sigma_{i,j}(A_{i,j}\hat{S}^i_x\hat{S}^j_x+B_{i,j}\hat{S}^i_y\hat{S}^j_y+C_{i,j}\hat{S}^i_z\hat{S}^j_z+D_{i,j}\{\hat{S}^i_x,\hat{S}^j_y\})+\Sigma_iH_i\hat{S}^i_z$, where $\{,\}$ is the anti-commutator. Here spin operators are defined in the standard way: $\hat{S}^i_z=1/2(|1\rangle_i\langle 1|_i-|2\rangle)_i\langle 2|_i)$, $\hat{S}^i_x=1/2(|1\rangle_i\langle 2|_i+|2\rangle)_i\langle 1|_i)$ and $\hat{S}^i_y=1/2i(|1\rangle_i\langle 2|_i-|2\rangle)_i\langle 1|_i)$. We have omitted terms proportional to $\hat{n}_i\hat{n}_j$, which are constant for pinned dipoles ($\hat{n}_i$, the molecular density is a substitute for the identity operator and is present in $H_i$). One unique feature of the current system is the asymmetry between the $A_{i,j}$ and $B_{i,j}$ exchange rates coming from the added/subtracted contribution of the spin-orbit part of the DDI $\propto\cos(2\phi_{ij})$ to the otherwise symmetric $A_{i,j}$ and  $B_{i,j}$ rates. The spin-orbit coupling adds also the anti-commutator term with a lattice index dependent rate $D_{i,j}\propto\sin(2\phi_{ij})$, which vanishes in 1D or a 2D square lattice. Since we envision a regime where both magnetic and electric DDI are of the same order $(|d_e|\tan(\theta)c^2/g_L\mu_B\sqrt{J(J+1)(2J+1)})^2\propto 1$ (neglecting Wigner 3-j coefficients ratios which are very similar in magnitude among the states of interest) one needs only a small admixture ($\theta_0\sim \pi/180$~rad). In this limit the corresponding rates can be expressed as $(A_{i,j};B_{i,j};C_{i,j};D_{i,j})=1/r^3_{i,j}(D_m+D_e\theta^2+3D_m\cos(2\phi_{i,j});D_m+D_e\theta^2-3D_m\cos(2\phi_{i,j});P_m+Q_e\theta^2;D_m\sin(2\phi_{i,j}))$ and the effective magnetic field is $H_i=\Sigma_j(R_m+S_e\theta^2)(1/r^3_{i,j})\hat{n}_{j}$, where we use the indices $e,m$ to refer to terms generated by the magnetic and electric DDI respectively and $\hat{n}_j=\Sigma_{\alpha=|1\rangle,|2\rangle}\hat{a}^{\dag}_{j\alpha}\hat{a}_{j\alpha}$ is the molecular density operator. 

One can vary the above rates by changing the mixing angle $\theta$ around $\theta_0$ which depends only on the ratio $\Omega/\delta$ and then adjusting the magnetic field to reach again degeneracy. On a typical optical lattice at 532~nm the above rates can be tuned in the vicinity of 50~Hz (the static mDDI rate for two atoms prepared in $|1\rangle$ is 150~Hz). Preparing neighboring degenerate states with lower angular momentum projection provides solely mDDI rates in the vicinity of 300~Hz. Moreover for such a choice of states, unlike the example discussed above, there will be also a spin-orbit eDDI. Therefore overall rates in the vicinity of 700~Hz are achievable with mixing angles of $\theta\sim 3\theta_0$. We expect a maximally mixed eigenvector in the degenerate basis, with ($\theta\sim\theta_0)$ to have a lifetime exceeding $250$~ms, limited by the theoretically estimated lifetime of the $|J=9\rangle$ state.  

A prerequisite for the success of our platform is a low magnetic noise environment. Such noise renders the two-level isospin system non-degenerate at all times. It can be modeled as a diagonal fluctuating term $\Delta_{ZN}$ in the full Hamiltonian matrix with standard deviation $\sigma_{ZN}=g_{10}\mu_B\sigma_{B}$ (we consider $g_{10}\approx g_9$ here, $\sigma_B$ is the standard deviation of the magnetic field noise). Therefore the magnetic field noise $\sigma_B$ should be on the order of a few $\mu$G. Experiments that demand such low magnetic field noise are typically isolated from the environment by a few layers of mu-metal~\cite{Arpaia2021, TheACMECollaboration2014}. The experimental complexity and the needed multi-beam optical access makes the current setup and in general cold-atom experimental setups hard to enclose in such mu-metal shields or makes such approach prohibitively expensive. Moreover the need for inner-shield coils for precisely setting the field demands their careful engineering that takes into account the non-linear influence of the high permeability nearby magnetic shield~\cite{Hobson2022}. Complex shields with multiple joints and access holes offer even lower suppression factors at AC stray magnetic fields~\cite{Dedman2007}. We therefore are developing a two-stage active noise cancellation system~\cite{Platzek1999May,Dedman2007,Merkel2019,Xu2019,Xiao2020Aug,Pyragius2021}. In the first stage we will simply use a feed-forward system, based on an arbitrary waveform generator, synchronized with the 50~Hz power line. The generator outputs a superposition of oscillating signals at the harmonics and the carrier of the power line and adds the correction to our current 3D set of large coils around the experiment, to counter the AC dominating noise. The cancellation is optimized once by minimizing the reading of a flux-gate magnetometer. The second stage of noise cancellation, already in operation, is based on a feedback system utilizing as an input the averaged error of two 3-axis fluxgate magnetometers symmetrically positioned with respect to the atomic sample (Bartington Mag-13 low noise option)~\cite{Xu2019}. A digital PID (Stefan Mayer Instruments, TWIN MR-3) processes the signal and sends the feedback correction to three sets of independent single loop rectangular coils with the size of 1.5$\times$2$\times 2$~m$^3$. The current scheme was thoroughly tested and brought the rms magnetic noise to $<100~\mu$G level. After the introduction of the additional feed-forward scheme we expect to be at a level of a few ppm, which should be confirmed ultimately with the atoms themselves~\cite{Xu2019}. 
\bibliography{newrefs_emil}

\end{document}